\documentclass[manuscript]{aastex}

\newcommand{\myemail}{mike.alexandersen (a) alumni.ubc.ca}

\newcommand{\unit}[1]{\, \mathrm{#1}} %Prints units nicely. 
\newcommand{\degree}{^{\circ}\!}
\usepackage{floatrow}
\usepackage{multirow}
\usepackage{hyperref}

\shorttitle{A characterised TNO survey}
\shortauthors{Alexandersen et al.}

\begin{document}

\title{A carefully characterised and tracked Trans-Neptunian survey, the size-distribution of the Plutinos and the number of Neptunian Trojans.}

\author{M. Alexandersen and B. Gladman}
\affil{Department of Physics and Astronomy, 6224 Agricultural Road, University of British Columbia, Vancouver, BC V6T 1Z1, Canada}
\email{\myemail}

\author{JJ. Kavelaars, Stephen Gwyn and Cory Shankman}
\affil{National Research Council of Canada, Victoria, BC V9E 2E7, Canada}

\and

\author{Jean-Marc Petit}
\affil{Institut UTINAM, CNRS-UMR 6213, Observatoire de Besan\c{c}on, BP 1615, 25010 Besan\c{c}on Cedex, France}

\begin{abstract}

The Trans-Neptunian Objects (TNOs) may preserve evidence of planet building in their orbital and size-distributions. 
While all populations show steep size-distributions for large objects, recently relative deficit of Neptunian Trojans and scattering objects with diameters $D<100\unit{km}$ were detected. 
We have investigated this deficit with a 32 square degree survey, detecting 77 TNOs to a limiting $r$-band magnitude of 24.6. 
Our Plutinos sample (18 objects in 3:2 mean motion resonance with Neptune) also shows a deficit of $D<100\unit{km}$ objects. 
We reject a single power-law size-distribution and find that the Plutinos favour a divot.
The Plutinos are thus added the list of populations with a deficit of $D<100\unit{km}$ objects. 
The fact that three independent samples of three different populations show this trend suggests that it is a real feature, possibly shared by all hot TNO populations as a remnant of ``born big'' planetesimal formation processes.  
We surmise the existence of $9000\pm3000$ Plutinos with $H_r\leq8.66$ and $37000^{+12000}_{-10000}$ Plutinos with $H_r\leq10.0$. 
Our survey also discovered one temporary Uranian Trojan, one temporary Neptunian Trojans and one stable Neptunian Trojan, from which we derive populations of $110^{+500}_{-100}$, $210^{+900}_{-200}$ and $150^{+600}_{-140}$ , respectively, with $H_r\leq10.0$. 
The Neptunian Trojans are thus less numerous than the main belt asteroids, which has over 700 asteroids with $H_r\leq10.0$. 
With such numbers, the temporary Neptunian Trojans cannot be previously stable Trojans that happen to be escaping the resonance now; they must be captured from another reservoir. 
With three 3:1 and one 4:1 resonators, we add to the growing evidence that the high-order resonances are more populated than expected.

\end{abstract}

\keywords{Trans-Neptunian Objects}

\section{Introduction} 

The outer Solar System (beyond $\sim 5\unit{AU}$) contains four giant planets, thousands of Jovian Trojans, at least four dwarf planets with diameters $D>1000\unit{km}$ and thousands of smaller TNOs whose numbers generally rise dramatically with decreasing size. 
The TNO populations have been shown to feature steep distributions for $D>100\unit{km}$ \citep{jewitt98,gladman01,bernstein04,elliot05,petit06,fuentesholman08,fraserkavelaars08}. 
For TNOs with $D>100\unit{km}$, the differential distribution of the Solar System absolute $H$-magnitude can be described well by a single exponential function, $dN/dH\propto10^{\alpha H}$, where $\alpha$ and the constant of proportionality are specific to the dynamical population of interest. 
Assuming a single albedo value for all objects and converting $H$-magnitude to diameter, the exponential magnitude-distribution is expressed as a power-law diameter-distribution $dN/dD\propto D^{-q}$, where $q=5\alpha+1$ is the differential size index\footnote{Note that while the cumulative diameter-distribution has a power law $\propto D^{1-q}$, the cumulative $H$-distribution retains the exponential exponent $\alpha$.}.
These steep distributions are thought to be imprints of planet accretion as $D<10\unit{km}$ objects accumulate into larger objects, with $q=4$ ($\alpha=0.6$), appearing in theoretical studies \citep{kenyonluu98,schlichting13}, while a steeper $q=5-7$ ($\alpha=0.8-1.2$) is commonly measured observationally for $D>100\unit{km}$ \citep{bernstein04,elliot05,petit11,adams14}. 
Other Solar System populations such as the asteroid belt and the Jovian Trojans have $q=4.5$ and $q=5.5\pm0.9$ for $D\gtrsim100\unit{km}$, respectively \citep{bottke05,jewitt00}, while $D<100\unit{km}$ main belt asteroids show $q=2.2$ to $4.1$ \citep{jedicke02,gladman09c}.

While the diameter-distribution is the real, physical property, we often do not know the albedo and thus do not know the size of observed objects, only their absolute magnitude; we therefore exclusively use the magnitude-distribution. 
Assuming a TNO $r$-band albedo of $5\%$ (as used by \citet{sheppardtrujillo10b,petit11,gladman12}, although this is in the low end of recently measured TNO albedos from \citet{mommert12,lacerda14}), $D\simeq100\unit{km}$ is equivalent to absolute $r$-band magnitude\footnote{Calculated using $H_r=m_{Sun,r}-2.512*log(0.05*(100\unit{km})^2/(4*(1\unit{AU})^2))$ with information from \url{http://mips.as.arizona.edu/~cnaw/sun.html}, \url{http://classic.sdss.org/dr4/algorithms/sdssUBVRITransform.html} and \url{http://www2.cadc-ccda.hia-iha.nrc-cnrc.gc.ca/en/megapipe/docs/filt.html}, giving $m_{Sun,r}=-27.09$.} $H_r=8.7$. 
The size-distribution of observationally accessible $100\unit{km}<D<1000\unit{km}$ TNOs have been the subject of intense study \citep{petit08,fuentesholman08}, while $D<100\unit{km}$ has only recently started to be investigated \citep{fraserkavelaars09}. 
Interpreting these results is complicated by the recent realization that there are multiple sub-populations present with different magnitude-distribution exponent \cite{bernstein04,elliot05,petit11,fraser10}.

Recent results claim a dramatic roll-over in the magnitude-distribution of Neptunian Trojans \citep{sheppardtrujillo10b} and scattering objects \citep{shankman13}, with a significant lack of $D<100\unit{km}$ objects. 
This might indicate that planetesimals were ``born big'' \citep{morbidelli09}, skipping the intermediate sizes, in sharp contrast with bottom-up formation \citep{kenyonluu98,weidenschilling97}. 
In the born big scenario, any intermediate sized objects we see today would be collisional fragments. 

Around a thousand TNOs are known\footnote{List of distant Solar System objects in the MPC database as of 2014 July 01: \url{http://web.archive.org/web/20140701182421/http://www.minorplanetcenter.net/iau/MPCORB/Distant.txt}}, however most of them were discovered in surveys that have not been ``well-characterised'' (as defined in \citet{jones06}) or have not had the characterisation published. 
Survey characterisations are important in order to perform accurate modelling of the biases of the observations. 
The Canada-France Ecliptic Plane Survey (CFEPS) is one of the largest surveys with published characterisation. 
However, CFEPS had very little sensitivity to objects with $D<100\unit{km}$. 
To improve upon the conclusions of CFEPS and to investigate the lack of $D<100\unit{km}$ objects described above, we performed a deeper survey designed to constrain the faint objects beyond the sensitivity limit of CFEPS. 

Several resonant populations may reach pericentre inside Neptune's orbit while still being long-term stable. 
Such TNOs can therefore be observed with even smaller diameters than those in the classical belt. 
Our survey was specifically designed to probe the Plutinos and the Neptunian Trojans, as the Plutinos are very numerous and come to pericentres near the Trojan clouds, within Neptune's orbit. 
We were able to probe beyond $D=100\unit{km}$ within these populations. 
Our goal was to confirm or refute the results in \citet{sheppardtrujillo10b}, which claimed a drastic drop in number density of Neptunian Trojans with $D<100\unit{km}$; we also hoped to investigate whether a similar paucity could be confirmed or rejected for the Plutinos. 
The Plutinos and Neptunian Trojans may have had a similar origin \citep{levison08,lykawka11}, so the Plutinos, being vastly more populous, would provide even stronger evidence for or against such a paucity. 

Results published after this survey was begun \citep{shankman13} suggested a similar sudden drop in the numbers of scattering objects with $D\lesssim100\unit{km}$ ($H_g\gtrsim9.0$), followed by a second power-law for faint objects, described as a ``divot'', similar to the divot postulated by \citet{fraser09c}. 
This result lends support to the idea that a drop in number density might be pressent in all dynamically hot TNO populations. 
However, \citet{fraser14} proposes that the dynamically hot TNOs share a broken power-law size-distribution (a ``knee'') which is initially steep but then breaks to a much shallower, although still rising, distribution. 
We have thus investigated single power-law, knee and divot scenarios, exploring which model parameters are compatible with our data.

Section \ref{sec:surveydesign} to \ref{sec:allobjects} describe the design and methodology of our survey, concluding with a summary of our detections. 
Readers uninterested in the technical details of the survey can skip to the analytic results of the Plutinos, Neptunian Trojans and Uranian Trojans in sections \ref{sec:plutinos}, \ref{sec:trojans} and \ref{sec:uranian}, respectively.

\section{Survey design}\label{sec:surveydesign}

\begin{figure}[htp]
\floatbox[{\capbeside\thisfloatsetup{capbesideposition={right,top}, capbesidewidth=7cm}}]{figure}[\FBwidth]
{\includegraphics[height=0.4\textheight]{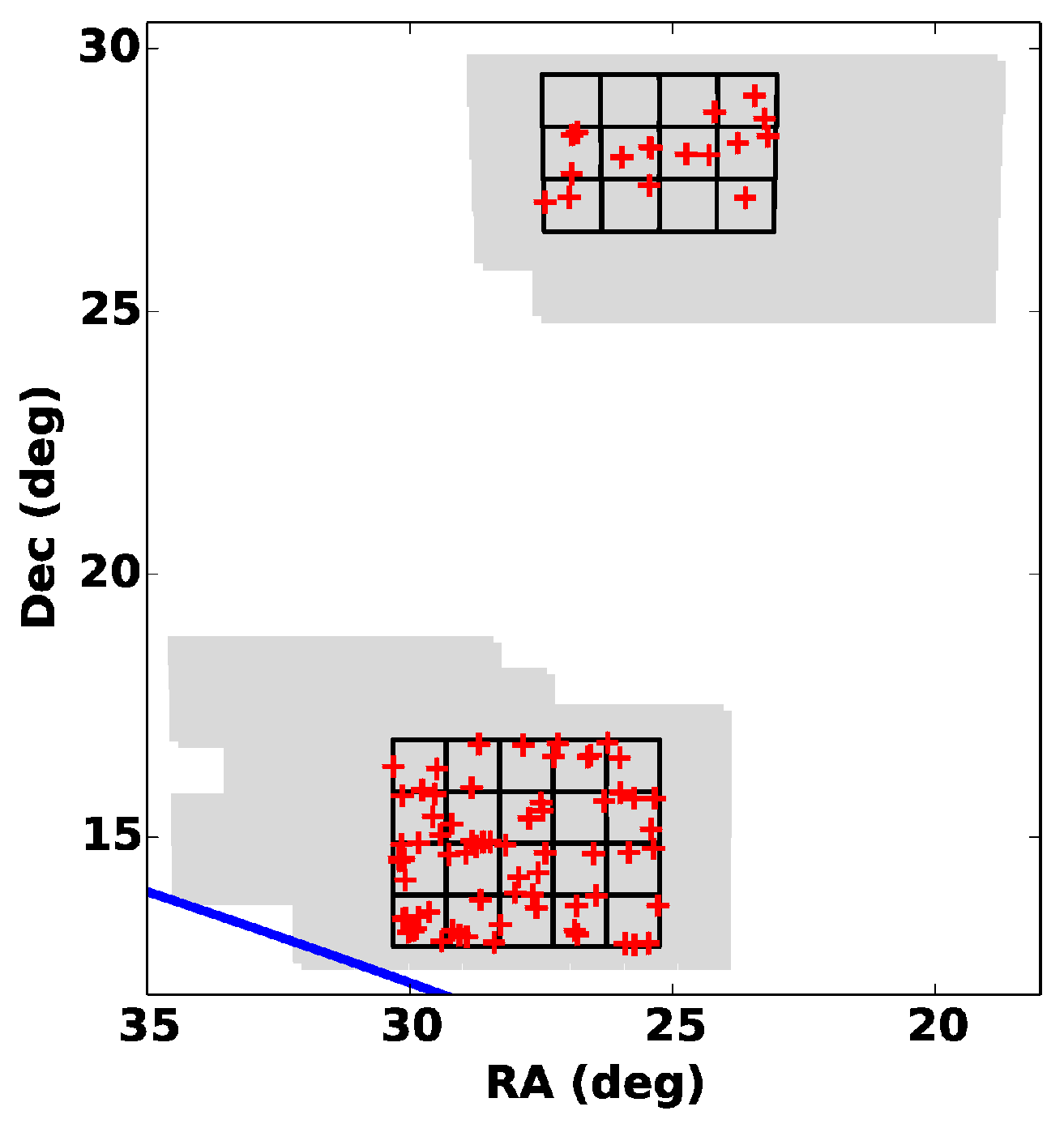}}
{\caption{\label{fig:coverage}
Sky coverage of our survey. 
Black boxes are the 1 sq.deg. field of view of CFHT's MegaCam, showing our discovery fields; 
crosses are the location of our detected objects at time of discovery; 
the inclined blue line in the lower left is the ecliptic. 
The shaded grey areas are the ``wallpaper'' coverage described in Sec. \ref{sec:wallpaper}.}}
\end{figure}

Our survey obtained 32 sq.deg. of high-cadence sky coverage using MegaCam on the Canada-France-Hawaii Telescope (CFHT), near RA=2hr, chosen to be near Neptune's L4 Lagrange point and near the pericentre point of many n:2, n:3, and n:4 resonances \citep{gladman12}. 
One `low-lat' block of 20 sq.deg. was centred near the ecliptic and another `high-lat' block of 12 sq.deg. was $\sim15\degree$ North (see Fig. \ref{fig:coverage} for block geometry and Tab. \ref{tab:pointings} for exact co-ordinates). 
The reason for dividing our survey into two blocks, locating a third of our survey well away from the ecliptic where the sky densitywould inevitably be lower, was to get a firm handle on the inclination-distribution of the populations. 
The Plutinos and Neptunian Trojans are dynamically hot populations, and thus their inclination-distribution is wide, with characteristic widths around $\sim15\degree$ \citep{gladman12,parker14}. 
High-inclination objects are strongly biased against being detected in ecliptic surveys, as these objects spend most of their time well away from the ecliptic; even with the drop in total number of detections, a survey at higher latitude should detect more high-inclination objects than and ecliptic one, providing a better handle on the shape of the inclination-distribution. 

Based on the sky density reported in \citet{sheppardtrujillo10b}, our survey would detect 3-4 Neptunian Trojans if the $D<100\unit{km}$ dearth was present, or 20-30 Neptunian Trojans if the lack of small objects in that work was a statistical fluke and the steep ($\alpha\approx0.8$) distribution continued. 
We thus chose our survey's magnitude limit to reach $m_r\approx24.5$ capable of detecting objects with $H_r\approx9.7$ at $30\unit{AU}$, more than a magnitude past the transition expected near $H_r\approx8.5\pm0.2$.

The discovery observations for each field were a set of three 320-s exposures separated by $\sim1\unit{hour}$ (``triplets'').
In the discovery year (2011 for low-lat, 2012 for high-lat), the discovery triplets were performed in October, close to opposition, when objects move faster and are slightly brighter. 
As reduction and characterisation of the observations and identification of objects takes several months, we could not perform targeted recoveries within the discovery year. 
However, as our blocks were large contiguous fields, shifting them across the sky at the mean TNO rate during the semester allowed us to successfully recover most objects in several other dark runs, with very limited loss due to Keplerian shear. 
For the low-lat block, our orbits were sufficiently well determined by the 1-opposition astrometry that we could perform targeted recoveries during the second year of observations, saving $\approx30\%$ of time compared to a contiguous block (while also having a higher recovery rate). 
As the high-lat block's discovery opposition was in 2012, we had ``precovery'' observations in 2011, consisting of a 20 sq.deg. contiguous shifted at the mean TNO rate. 
This approach successfully precovered all but one nearby, highly inclined Neptunian Trojan. 
This object, however, was observed again in 2013, along with most of our Plutinos.  

\subsection{Observations}
The discovery images were taken in roughly three-hour stretches, with exposure times of $320\unit{s}$. 
MegaCam at CFHT has an overhead time of $40\unit{s/image}$, resulting in a 10 exposures per hour imaging rate. 

The ''low-lat'' block ($5\degree$x$4\degree$ fields centred on 01:51:08 +14:53:00) was divided horizontally into two 10 sq.deg. sub-blocks. 
The southern half obtained discovery observations on 2011/10/24, the northern half on 2011/10/26. 

The ''high-lat'' block (4x3 fields centred on 01:45:30 +28:10:00) was divided vertically into two 6 sq.deg. sub-blocks. 
The discovery triplets consisted of the 6 fields of the sub-block, interspersed with single images (``nailing'' images) of fields from the other sub-block to extend the interval between subsequent images of the triplets up to near an hour. 
The western half obtained discovery observations on 2012/10/20; the eastern half was observed on 2012/10/21. 

Objects were tracked for up to 28 months of total arc. 
Single ``nailing'' images were taken at various times to extend the arc of the astrometry. 
At least one nailing image is required in the discovery dark run; two is preferable, in order to prevent losing objects in chip-gaps. 
Once an object has been located on another night in the discovery dark run (extending its arc to a few days), it can usually be located in images from one month before or after discovery. 
This iterative process continued until the object has been found on all available images, or the remaining images have been deemed unusable.  
By the end of the discovery year, most objects had several months of well-sampled arcs, so the on-sky predicted locations for the following/previous years were accurate to a few arcseconds; for the low-lat block, pointed recoveries were thus trivial. 
One limitation in extending the arc of observations with single nailings is that you must have a comparison image in order to distinguish moving objects from stars, so most dark runs had two nailings of each field, taken on separate nights. 

\subsection{Accurate astrometry}\label{sec:wallpaper}

The uncertainty in the heliocentric orbits of a TNO depends on the observed arc, the number and temporal distribution of observations inside the arc, and the accuracy of the measured astrometry. 
Larger astrometric uncertainty allows a wide range of acceptable orbital fits, and astrometry affected by systematic errors push the determined orbit away from the true parameters. 
Regular recovery over long arcs can help counter these problems, but obtainng the best-possible orbit-determinations in a short amount of time requires the best possible astrometry. 
A particular concern with short arcs is systematic errors, which can occur when an object moves from one region to another, where the astrometric references might have systematic offsets of order $\sim0.2''$.
To limit systematic errors in our astrometry as much as possible, two large plate solutions and stellar catalogues were made (one for each block), which served as block-wide astrometric reference catalogues. 

To make the unified astrometric catalogue, every image from 2011-2012 was stitched together. 
The slow drift of fields across the sky, produced considerable overlap, allowing the fields to be stitched together and to fill in chip-gaps. 
To ensure that all portions of the sky coverage were connected (including outlying fields from pointed recoveries in 2012), we also shot two grids of ``wallpaper''; these were shallow, 20-second exposures in two offset grids covering all our other pointings. 
The two grids were offset by half a chip in the vertical direction and 1.5 chips in the horizontal direction to ensure overlap between neighbouring fields and to remove the distortion of the focal-plane. 
The coverage of the wallpaper, and thus the extent of our astrometric catalogue, can be seen in Fig. \ref{fig:coverage}.

The astrometry of each MegaCam image, both the science image and the wallpaper images, was individually calibrated using the 2MASS catalogue \citep{skrutskie06} as a reference. 
Catalogues containing RA, Dec and magnitude were created for each image. 
These catalogues were merged into a single catalogue: sources found in multiple images were identified and their positions averaged together. 
This merged catalogue was used to calibrate all the images. 
The resulting astrometric calibration has internal uncertainties on the order of 0.04 arcseconds.

\begin{figure}[htp]
\floatbox[{\capbeside\thisfloatsetup{capbesideposition={right,top}, capbesidewidth=0.49\textwidth}}]{figure}[\FBwidth]
{\includegraphics[width=0.49\textwidth]{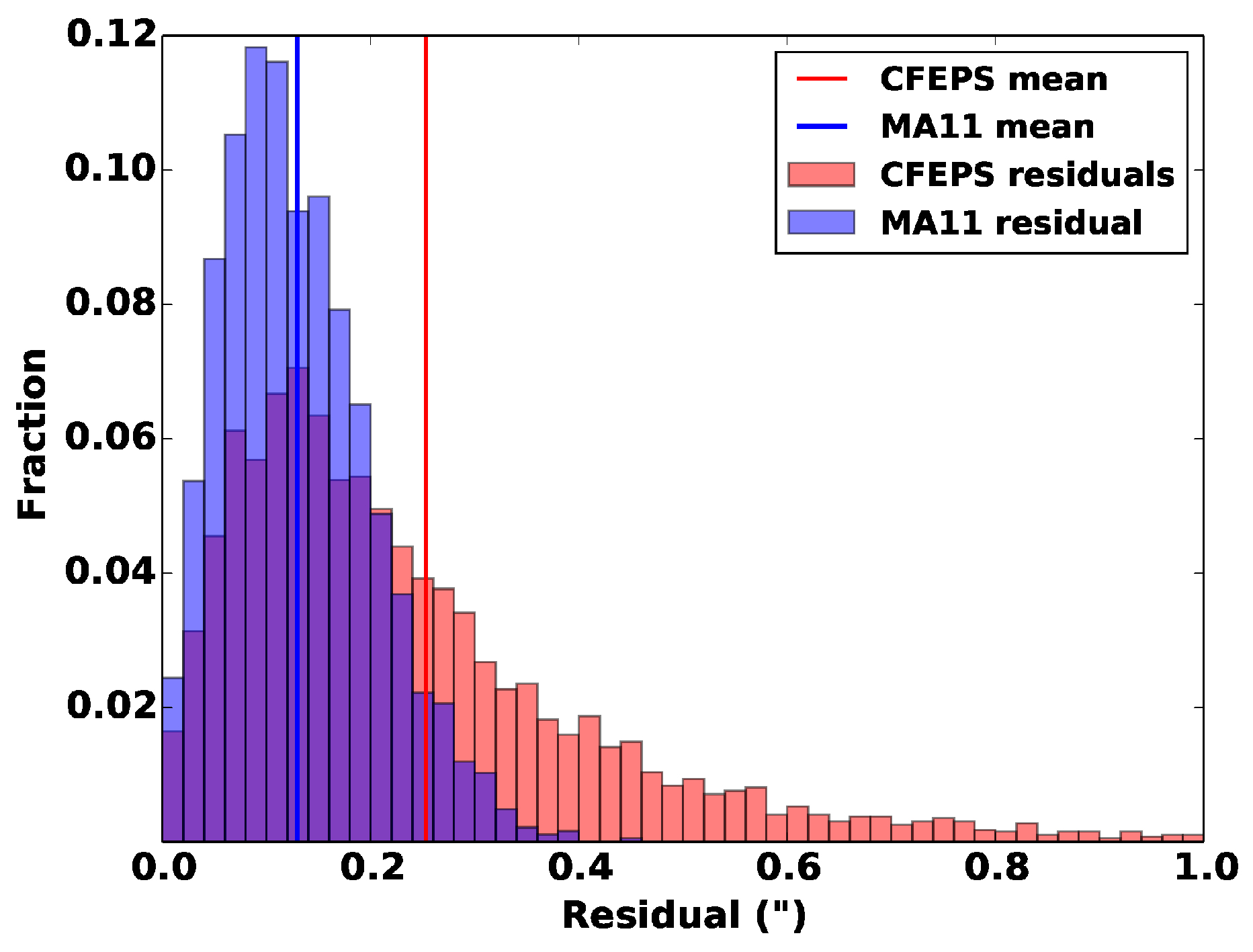}}
{\caption{\label{fig:residuals}
Histogram of all residuals for orbit-fits to astrometry from CFEPS (red) and this survey (MA11, blue), with the survey means shown with vertical lines.
Residuals are the x and y residuals combined in quadrature.  
}}
\end{figure}

This survey-wide astrometric catalogue allowed us to measure accurate astrometry over a two year period leading to unprecedented small residuals from our orbit-fits, with mean residuals of $0.13''$ and maximum residual of $0.45''$ (where x and y residuals have been combined in quadrature). 
These residuals are dominated by measurement errors in the positions of the objects themselves (due to low signal to noise ratio), not by uncertainties in the astrometric catalogue. 
This is a vast improvement over the $0.25''$ RMS residuals of CFEPS (see Fig. \ref{fig:residuals}), providing better orbit-determinations with a short arc. 
This improvement allowed us to perform pointed recovery observations of almost all low-lat objects in 2012B, with just six-month arcs in 2011B, and to securely classify most objects at the end of 2012B, see Sec. \ref{sec:orbits}.

\section{Characterisation Methods}

The triplets from the discovery night were passed through the moving object pipeline developed by the CFEPS team \citep{petit04}. 
This pipeline employs two different moving-object detection algorithms to search for linearly moving sources.  
 To minimise the number of false candidates (where the pipeline detects 
 background noise or cosmic rays), only the overlapping set of detections
 shared by the two algorithms are considered ``candidate'' objects; 
these candidates were subsequently vetted by human inspection. 

In order to determine the characteristics of our survey (detection-efficiency and false positive rate) we also planted artificial objects using methods outlined in \citet{petit04} and \citet{jones06}. 
We here present a new improvement to this process; we did not plant the artificial objects into the ``real'' triplets that was searched for real objects. 
Instead we used a copy of the triplet images, in which we had first temporally scrambled the order of the images. 
That is, we switched around the UT timestamps in the image headers so that images 1, 2, 3 were permutated to the sequence 2, 3, 1. 
This scrambling of order meant that any detections by the pipeline in these ``fake'' image triplets must either be artificially-implanted objects or be false positives; nothing can be real objects, because no real object could move linearly in these out-of-order images. 
That is, no real outer Solar System object can reverse apparent direction of sky motion within two hours.  
To our knowledge, this approach, which allows us to measure the false positive rate exactly, has never been used before. 
This type of science (with short-term monotonic time variability) is uniquely able to perform such false positive measurements, and we propose that this method becomes standard for measuring false positive rates. 

Once the real and fake triplets had been passed through the automated pipeline, a human operator inspected every candidate from both sets of images simultaneously. 
The candidates from the two sets of images were mixed together, such that the operator never knew whether a candidate from a real or fake triplet is being shown. 
The blind inspection ensured that the two sets were given the exact same treatment. 

During the human inspection, detections which were not believed to be valid (mostly caused by noise and stellar diffraction spikes) were rejected by the operator. 
The operator also assigned flags to observations, flagging images where the astrometry or photometry might not be trustworthy. 
This included images with stellar interference in the object's point spread function or bright sources significantly polluting the photometric sky annulus.

Once all candidates had been inspected, having been either approved or rejected, the approved objects in the fake set of triplets provided the efficiency function and the false positive rate. 
The approved objects in the real set of triplets are real objects (or false positives) and were subsequently hunted for in tracking observations. 

\subsection{Efficiency of search}
The detection efficiency was calculated as a function of magnitude, for a few rate of motion ranges, in order to investigate whether our efficiency varied significantly with rate. 
These efficiencies as a function of magnitude was found to be well represented by a function described in \citet{gladman09c}, of the form: 
\begin{equation}\label{eq:eff}
f(m_r)=\frac{f_{21}-k(m_r-21)^2}{1+\exp\left(\frac{m_r-m_e}{w}\right)},
\end{equation}
where $f_{21}$ is the efficiency at $m_r\simeq21$, $k$ is a measure for the strength of a quadratic drop (caused by crowding; objects being obscured by background stars is a problem that increases for faint objects), $m_e$ is the magnitude at which the function transitions to being an exponential tail and $w$ is a measure for the width of that tail.

\begin{figure}[htp]
\includegraphics[width=1.00\textwidth]{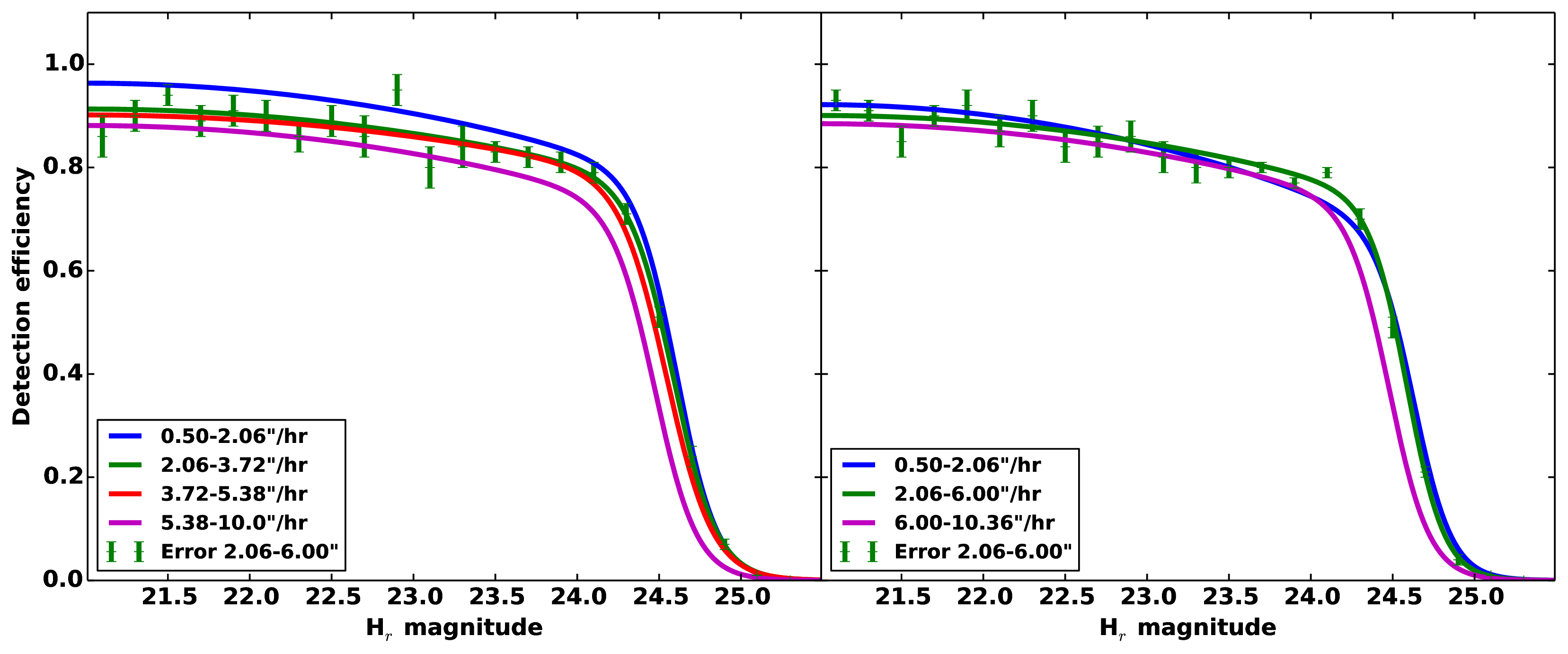}
\caption{\label{fig:efficiency}
Efficiency functions for our survey, low-lat block (left) and high-lat block (right). 
To prevent crowding, error-bars are shown only on the curve most applicable for TNOs at $\sim40\unit{AU}$ is shown.
}
\end{figure}

The limiting magnitude of the survey for each rate range was chosen to be the magnitude at which the efficiency was $40\%$ of the maximum efficiency, rounded up to nearest $0.01$ magnitude. 
The limiting magnitudes for the blocks are given in Tab. \ref{tab:limits}, as well as the parameters used to model the efficiency functions, which are shown graphically in Fig. \ref{fig:efficiency}. 
As all but two of our objects, the ones detected at $r<28\unit{AU}$, have rates between $2.5$ and $4.4\unit{"/hr}$, we can see that the survey limit for main-belt TNOs is $m_r\simeq24.6$. 

\subsection{False positive rate}

The fake triplets produced had 22910 candidate detections, of which 1052 were rejected and 21858 were accepted; of the accepted objects 21855 were planted, leaving 3 false positives. 
In other words, the human operator was willing to accept these three detections as real candidates near the noise limit, which were actually coincidental noise. 
Having only three false positives despite the sheer number of candidates shows the ability of humans to successfully filter noise from valid low signal-to-noise sources. 
Moreover, these three false positives were all beyond our survey's characterisation limit, so in fact the characterised portion of our survey had a zero false positive rate.
This is the first time such a measurement of false positive rate has been made, and suggest that it should become standard. 
Our result of zero false positives in the characterised portion of the survey should not be seen as evidence that surveys like this always have a zero/low false positive rate, but rather serves as evidence of the effort we put into the candidate inspection process. 
If a significant number of false positives had appeared that were brighter than the characterisation limit, it would call into question the validity of the efficiency function and thus one's ability to de-bias the survey. 

In the real triplets, there were 1246 candidates; 1159 of these were rejected, leaving 87 candidate real objects, 70 in the low-lat block and 17 in the high-lat block. 
These objects were searched for in tracking observations in 2011-2013. 

\subsection{Tracking}

We were able to track almost every object detected in our survey. 
Ten objects did not achieve two-opposition arcs, however, eight of these were past our characterisation limit and thus do not affect our scientific modelling. 
The loss of the two objects brighter than our characterisation limit (with mags $m_r=24.2$ and $24.5$) was due to a combination of poor cadence and image quality of some tracking observations. 
When the nailing images within the discovery dark-run are not sufficiently deep to see the object, it is very hard to find the object in images several months away, as the error-ellipse can be wider than the CCD chip, $\sim7'$. 
Because the characterization indicated that we have a zero false-positive rate in our characterised survey, we believe that these are two real objects. 

To account for the (slightly) less than $100\%$ tracking efficiency in our analysis in the coming sections, we modelled the tracking efficiency based on the loss of these two objects, one in each block. 
We do not believe that our tracking efficiency is dependant on the orbit of the object, as our observations within the discovery year were always a large contiguous patch of sky, meaning tracking did not rely on assumed orbits.
The low-lat untracked object, mal11nt = 2011UU$_{412}$ with $m_r=24.5\pm0.1$, is near our characterisation limit and was simply invisible in the slightly shallower tracking images; we therefore modelled the tracking efficiency of the low-lat block as $100\%$ for $m_r<24.5$ followed by a linear falloff such that integrating over the tracking function gives the appropriate fraction of lost objects. 
In the high-lat block, mah11nt with $m_r=24.2\pm0.1$ was not succesfully tracked, despite being a fair bit brighter than our limit. 
The high-lat block lost mah11nt (no MPC designation due to short arc) due to poorer observing cadence and the fact that mah11nt sheared off the field during part of the discovery year; its lack of recovery is thus independent of the object's brightness, so we modelled the tracking efficiency of the high-lat block as a constant $93\%$. 
These tracking efficiencies were incorporated into our survey simulations.

\section{Orbit classification}\label{sec:orbits}

At the end of 2012B, our objects typically had 17 month arcs, with 15-30 observations in multiple months in both 2011 and 2012. 
The cadence of observations together with the superior astrometry ensured that with just two oppositions, most of our objects have their orbits determined to far better precision than most two opposition objects in the MPC database; those objects often have the majority or entirety of their observations in the opposition months. 
In order to classify objects with reasonable confidence, one needs $\sigma_a/a\lesssim0.003$, and in order to constrain the libration amplitude of resonant objects, $\sigma_a/a\lesssim0.001$ is typically required. 
71 and 59 of our objects had $\sigma_a/a$ smaller than these values, respectively, by the end of 2012B. 

The objects were dynamically classified using the Solar System Beyond Neptune nomenclature \citep{gladman08}. 
The nominal orbit, as well as the two orbits with the most extremal semi-major axis allowed by the astrometry, were integrated for $10\unit{Myr}$. 
The two extremal orbits are determined from a Monte-Carlo process. 
If integrated particles started on these three orbits exhibit the same behaviour (resonant, classical, detached, scattering, etc.), the object is said to be securely classified, but if all three clones do not experience similar behaviour, the object is insecurely classified. 
Tab. \ref{tab:objects} list the resulting orbital classes and security of the classification.  

\label{sec:allobjects}

In our discovery triplets, 87 objects were detected; 
 70 in the low-lat block and 17 in the high-lat block. 
 77 of these 87 were above our characterisation limit, comprising our characterised sample. 
Among the characterised sample, we found 
 one Uranian Trojan (1:1 mean-motion resonance with Uranus, see \citet{alexandersen13b}), 
 two Neptunian Trojans (1:1 resonance with Neptune), 
 two 4:3 resonant objects,
 18 Plutinos (3:2 outer resonance with Neptune), 
 six 5:3 objects, 
 three 2:1 objects, 
 two 5:2 objects, 
 three 3:1 objects, 
 one 4:1 object
 and a slew of main belt and detached objects. 
We present the entire sample, with orbital information and classification, in Tab. \ref{tab:objects}. 
However, from here on we will only use the characterised sample, focusing on the Plutinos in Sec. \ref{sec:plutinos}, Neptunian Trojans in Sec. \ref{sec:trojans}, Uranian Trojans in Sec. \ref{sec:uranian} and the 4:1 resonance in Sec. \ref{sec:fourtinos}.  

\section{Plutino analysis}\label{sec:plutinos}

We have analysed the orbital and magnitude-distribution of the Plutinos in detail. 
Our survey detected 18 Plutinos, with eccentricity, inclination, discovery distance, apparent and absolute magnitude in the ranges $e=(0.04,0.33)$, $i=(2.7\degree,23.8\degree)$, $r=(28.7,45.5)\unit{AU}$, $m_r=(21.5,24.5)$, $H_r=(5.5,9.9)$, respectively. 
For much of the following analysis, we used a Survey Simulator nearly identical to that developed by the CFEPS team \citep{kavelaars09}. 
Given a model distribution of TNOs, objects with orbital-parameters and $H$-magnitudes were generated. 
The synthetic objects were then passed to the survey simulator which exposed them to the biases of the characterised survey (field location, efficiency function, limiting magnitude, tracking fraction, etc), determining which of the the synthetic objects would have been detected and tracked to high-quality orbits. 
Subsequently, the simulated detections and real detections can be compared, providing a test of the model distribution's validity. 

  \subsection{Statistical method}\label{sec:stats}

Throughout this work, except Section \ref{sec:twobin}, we have used a boot-strapped four-parameter Anderson-Darling test when comparing real detections to simulated ones. 
The Anderson-Darling (AD) statistic \citep{andersondarling54} was calculated for the real object set compared to simulated detections (containing ten thousand to fifty thousand simulated detections) for four properties: eccentricity, inclination, heliocentric distance at discovery and absolute magnitude. 
These parameters were because they best reveal significant shortcomings of a combined orbit and magnitude-distribution model; for example, as all Plutinos have semi-major axes values within $1\%$ of each other, detailed modeling of the $a$-distribution is unimportant, as small $a$ variations in this range have no discernable effect on detectibility.  
The sum of the four statistic was saved, to be compared to bootstrapped values.
The sum is used\footnote{Unlike, for example, the Kolmogorov–Smirnov test, the AD test can be added in multiple dimensions like this as its statistic is a unitless, normalised quantity.}, as it will be small if all four distributions are in good agreement, large if one of the distributions is a poor match and even larger if several distributions are poor matches \citep{parker14}.
Subsamples with the same number of detections as the real set were drawn from the simulated detections, and their summed statistic was calculated in the same way; this process was repeated one thousand times. 
The distribution of summed AD statistics for the simulated subsets reveal the probability of a random subset being a worse match to the parent simulated detections than the real set is, giving the probability that the real set could be drawn from the simulated detections. 
Probabilities $<5\%$ or $<1\%$ are considered thresholds of significant or highly significant rejection, respectively. 

  \subsection{Comparison with CFEPS L7 model}

The magnitude limit of CFEPS was more than a magnitude brighter than our survey's, but CFEPS had a much larger area, resulting in 24 Plutino detections. 
The CFEPS L7 model of the Plutino size- and orbital-distributions was constructed based on these 24 Plutinos \citep{gladman12}.
We thus first investigated whether our independent Plutino sample was in agreement with the L7 model. 
Using our set of 18 Plutinos, we found that our Plutinos reject the L7 model if the $H$-distribution is extrapolated to $H_r=11$, at $98\%$ confidence. 
That is, $P<2\%$ for drawing the real detections from the model's simulated detections, with the $H$-magnitude distribution clearly appearing to be the culprit. 
Truncating the detections (real and simulated) at $H_r=8.66$ (the magnitude of the faintest CFEPS Plutino), our 12 remaining Plutinos still reject the L7 model at $98\%$ confidence. 
It is interesting to note that our survey has no Plutino detections with $8.27<H_r<9.01$, yet found six Plutinos (a third of the sample) with $H_r\geq9.01$. 
Truncating the detections at $H_r=8.3$, our remaining 12 Plutinos suddenly provided a non-rejectable match with the model, with 21\% probability that the observations could be drawn from the model.  
This suggests that the problem is indeed that the magnitude-distribution does not extend with the same exponential form to fainter magnitudes, with a drastic change in the magnitude-distribution around $H_r=8.3$. 
This $H_r\simeq8.3$ transition is very similar to the drop in number density at $H_R=8.5$ ($H_r=0.7$) found for the Neptunian Trojans \citep{sheppardtrujillo10b} and the $H_g=9.0$ ($H_r=8.5\pm0.2$) transition concluded for the Scattering Objects \citep{shankman13}, as we expected it would be when we designed the survey. 
This commonality is explored further later in this manuscript. 

  \subsection{Improved orbital and magnitude-distribution model} \label{sec:bestmodel}

From this point onwards, the combined detections and survey characterisation of CFEPS and our new survey was used.  
As our survey was performed in $r$-band and CFEPS was primarily performed in $g$-band, an assumption of colour had to be made when combining the surveys. 
Most CFEPS Plutinos had their $g-r$ colours measured \citep{petit11}; the few that did not have measured colours had their magnitude converted using the average value of the CFEPS Plutinos with measured colours ($g-r=0.5$). 
In the Survey Simulator, the two surveys were combined assuming a $g-r$ colour-distribution based on the colour-distribution of the CFEPS Plutinos, modelled as a Gaussian distribution of width $0.2$ centred on $0.5$. 

\begin{figure}[tbph!]
\includegraphics[width=0.85\textwidth]{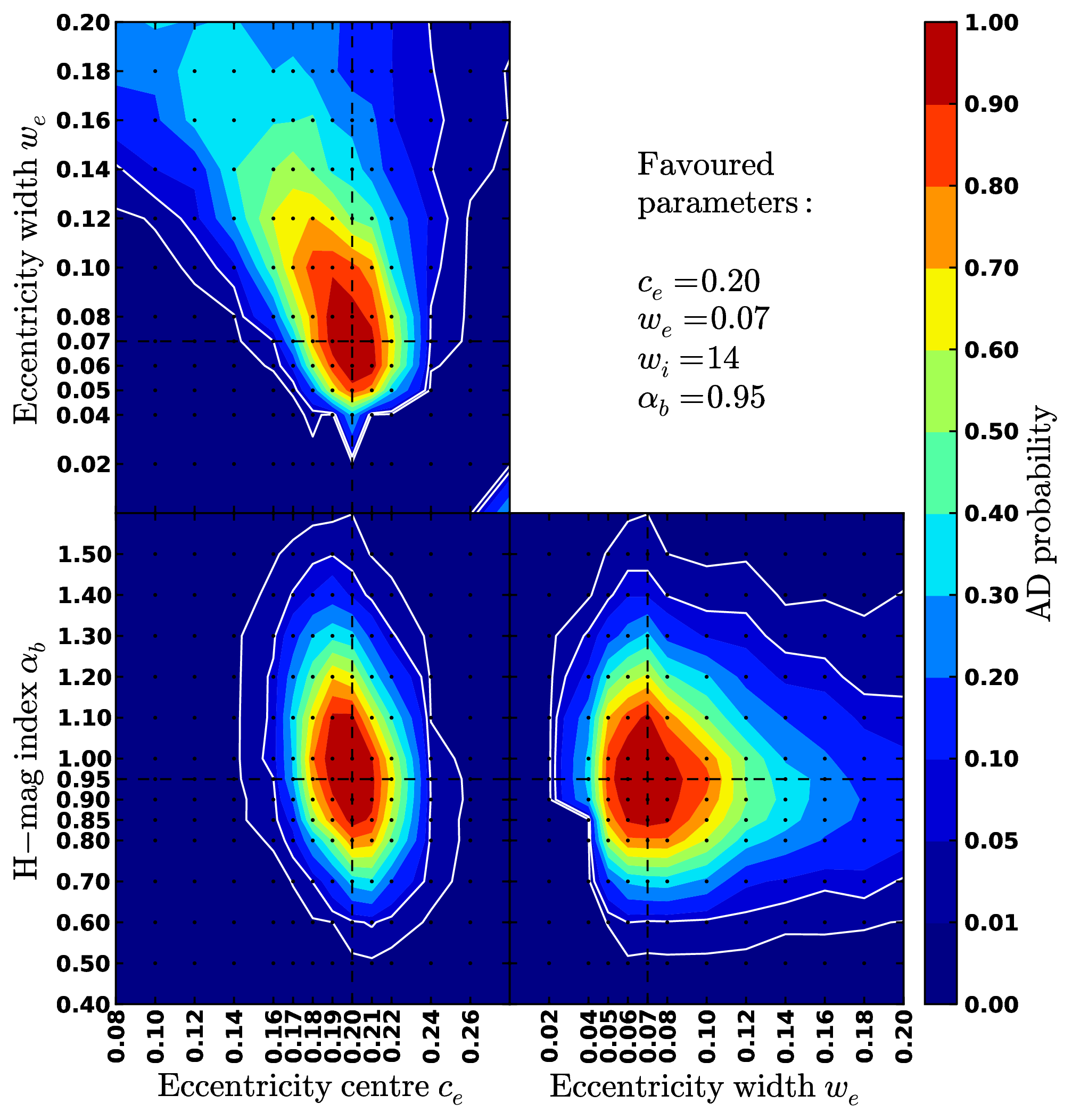}
\caption{\label{fig:contour-bestorbit}
Contour plots of a parameter space search, for $H_r\leq8.3$. 
This shows the AD probability of our real detections with $H_r\leq8.3$ against simulated detections from models with a grid of $c_e$, $w_e$, $w_i$ and $\alpha_b$ parameters run through the survey simulator. 
This plot is equivalent to Fig. 4 in \citet{gladman12} and the patterns are in good agreement, with a peak only slightly offset, by values smaller than the uncertainties. 
Contours are only shown for $w_i=14\degree$, as the shape of the contours are similar for all $w_i$, but the values peak at $w_i=14\degree$. 
The three plots are slices through the three-dimensional space, such that each plot is fixed to the best value of one parameter while the other two are varied. 
The full four-dimensional grid was integrated (including a range of $w_i$ values) in order to find the best solution. 
White lines outline the $1\%$ and $5\%$ contours in the planes.  
}
\end{figure}

To improve our Plutino model, we first refined the model to be the best match for objects brightward of the putative transition. 
We continued using the same parameterization as in CFEPS\footnote{
The only change we made to the model was for the libration amplitude, correcting a coding-error in the CFEPS routine; rather than having a triangle starting at $20\degree$ as described, the old code had a triangle starting at $0\degree$ which had values $<20\degree$ truncated off. Our change led to slightly fewer low-amplitude objects being generated, however, this change did not lead to any large change in our results. 
}, but wished to improve the parameter values. 
The parameters we wished to improve were the width $w_e$ and centre $c_e$ for the Gaussian eccentricity-distribution ($\mathrm{e}^{\frac{(e-c_e)^2}{2w_e^2}}$), the width $w_i$ of the Brownian inclination-distribution ($\sin(i)\mathrm{e}^{\frac{i^2}{2w_i^2}}$) and the exponent $\alpha_b$ of the exponential $H$-magnitude distribution ($10^{\alpha_b H_r}$), where the subscript denotes that we were only investigating the bright objects (brighter than the transition). 
These model distributions have no cosmogonic motivation, but CFEPS found them to work well. 
Given that these models, with the right parameters, continue to be able to achieve a high AD-probability (see Fig. \ref{fig:contour-bestorbit}), their functional forms are still appropriate.

We ran survey simulations with models using a four-dimensional grid of the $c_e$, $w_e$, $w_i$, $\alpha_b$ parameters, for Plutinos with $H_r<8.3$, using a single exponential magnitude-distribution, $10^{\alpha_b H_r}$. 
This $H_r$ cut was chosen in order to only include objects brightwards of the transition, while not removing so many objects that it degrades our statistics. 
CFEPS and our new survey combined has 34 Plutinos with $H_r<8.3$.

\begin{figure}[tbp]
\includegraphics[width=1\textwidth]{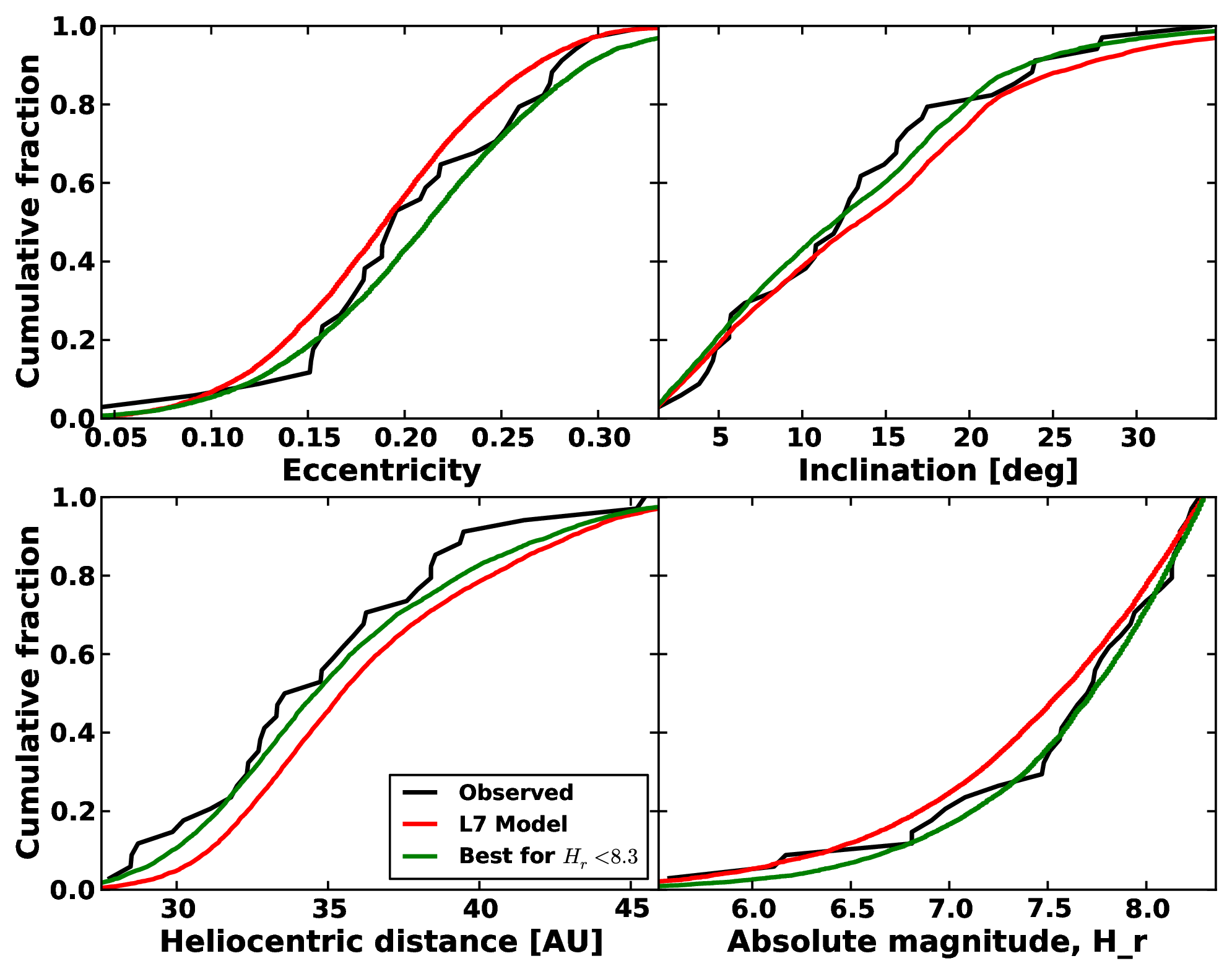}
\caption{\label{fig:cumul-bestorbit}
Cumulative distributions of the eccentricity, inclination, heliocentric distance at discovery and absolute magnitude of real (black) and simulated detections, for the $H_r<8.3$ sample. 
The favoured model (green) has $c_e=0.20$, $w_e=0.07$, $w_i=14\degree$, $\alpha_b=0.95$, while the L7 model (red) has $c_e=0.18$, $w_e=0.06$, $w_i=16\degree$, $\alpha_b=0.90$.
Raising $c_e$, $w_e$ and $\alpha_b$ all contributed to the detection of more small objects at the closest distances. 
}
\end{figure}

From this analysis, we found that only minor adjustments could be made to the L7 model parameters to improve its agreement with observations, the largest change being that the $H_r<8.3$ sample favoured a slightly smaller inclination width of $w_i=14\degree^{+8\degree}_{-4\degree}$ althought this is still well within the CFEPS range of $16\degree^{+8\degree}_{-4\degree}$ \citep{gladman12} (uncertainties represent ranges for which the propability $>5\%$, when all other parameters are fixed). 
The model's other parameters were also tweaked slightly, with our favoured parameters now being $c_e=0.20^{+0.02}_{-0.04}$, $w_e=0.07^{+0.15}_{-0.03}$, $w_i=14\degree^{+8\degree}_{-4\degree}$, $\alpha_b=0.95^{+0.45}_{-0.25}$ (the previous L7 parameters being $c_e=0.18$, $w_e=0.06$, $w_i=16\degree$, $\alpha_b=0.90$). 
This slightly steeper slope of $\alpha_b=0.95^{+0.45}_{-0.35}$ ($95\%$ confidence), constrained using the range $5.5<H_r<8.3$, is comparable to recent results analyses of the Deep Ecliptic Survey \citep{adams14}, which measured $\alpha_b=0.95\pm0.16$ ($1\sigma$) over the range $4.6<H_r<7.1$. 
See Fig. \ref{fig:contour-bestorbit} for a contour plot showing the acceptable range of parameters, and Fig. \ref{fig:cumul-bestorbit} for a visual representation of the simulated and real detections illustrating the improvement gained from tweaking the parameters. 
An important correlation remains that a lower-eccentricity peak for the centre $c_e$ of a Gaussian eccentricity-distribution requires a larger width $w_e$ to produce the copious numbers of $e>0.2$ Plutinos. 

\begin{figure}[tbp]
\includegraphics[width=0.75\textwidth]{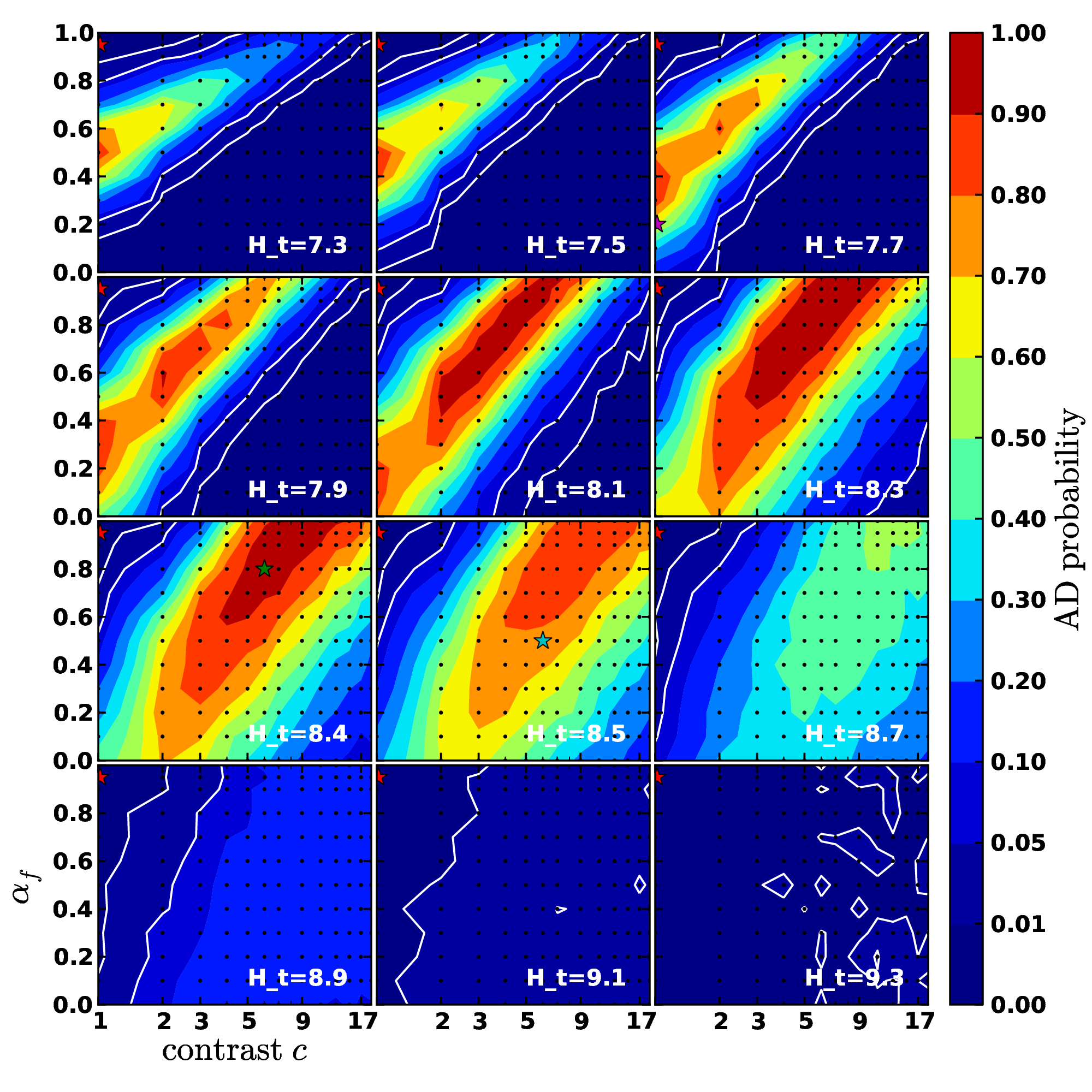}
\caption{\label{fig:contour-bestdivot}
Contour plots of our $H_t$, $c$, $\alpha_f$ parameter space search, with $\alpha_b=0.95$ and orbital parameters as derived in Sec \ref{sec:bestmodel}. 
Each subplot has the same axes and is a slice through the three-dimensional space, sliced at various values of $H_t$ (labelled on each subplot). 
White lines outline the $1\%$ and $5\%$ probability contours. 
With $H_t>9$, all solutions are rejectable, because the steep $\alpha_b$ cannot continue to $H_r=9$; there must be a transition at a brighter magnitude. 
In the bright end, knees (solutions with $c=1$) are allowable. 
However, the least rejectable models are divots around $H_t=8.4$. 
Note that knees to $\alpha_f\geq0.8$ are always rejectable at $95\%$ confidence, and for $H_t>8.2$, knees to $\alpha_f\geq0.6$ are also rejectable. 
Our favoured parameters are marked with a green star on the $H_t=8.4$ panel. 
Also marked are the knee parameters favoured by \cite{fraser14} for the hot TNOs (magenta star on the $H_t=7.7$ panel), the divot parameters favoured by \citet{shankman13} for the Scattering TNOs (cyan star on the $H_t=8.5$ panel) and the single exponential model (red star on every plot, as varying $H_t$ when $c=1$ and $\alpha_f=\alpha_b$ changes nothing). 
}
\end{figure}

\begin{figure}[tbp]
\includegraphics[width=1\textwidth]{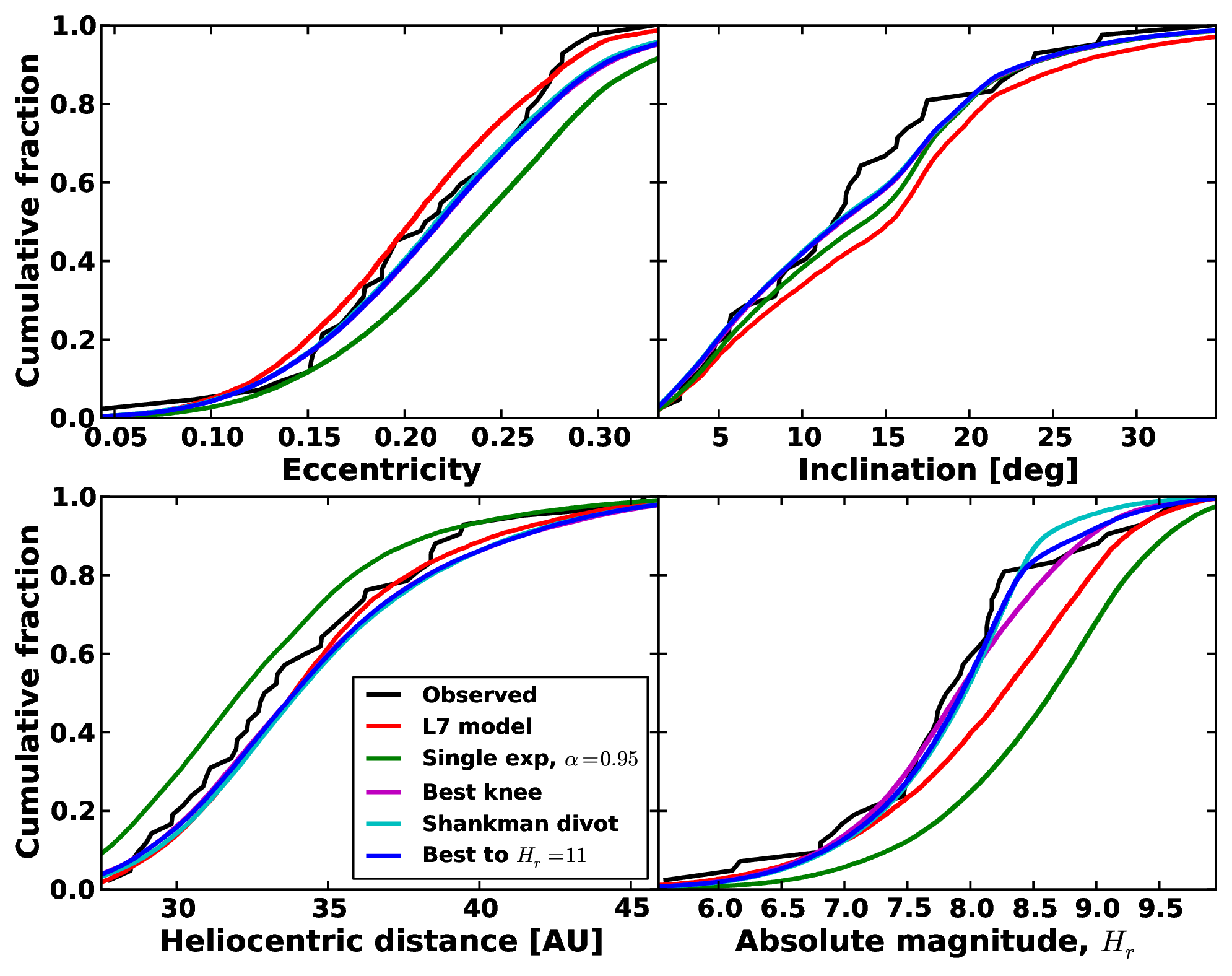}
\caption{\label{fig:cumul-bestdivot}
Cumulative distributions as in Fig. \ref{fig:cumul-bestorbit}, for the full combined Plutino sample. 
Our simulations generated objects down to $H_r=11$, well beyond our faintest real detection of $9.85$, to prevent artificially cutting the simulated detections. 
The favoured solution (dark blue) has $H_t=8.4$, $c=6$, $\alpha_f=0.80$, while the best knee (magenta) had $H_t=7.7$, $c=1$, $\alpha_f=0.40$. 
It is clear that both of these solutions provide a better representation of the observations than the single exponential of same $\alpha_b$ (green) and the L7 model (red) over this full magnitude range (compare to Fig. \ref{fig:cumul-bestdivot}). 
For comparison we also shown the divot with $H_t=8.5$, $c=6$, $\alpha=0.5$ (cyan) as found for the Scattering Objects \citep{shankman13}. 
Note that the cyan, magenta and blue curves are almost identical and difficult to distinguish in the two top panels and bottom left panel. 
}
\end{figure}

Making the reasonable assumption that the orbital-distribution is not $H$-mag dependent, we fixed the Plutino orbital-distribution to the parameters found above.
This allowed us to explore the remaining $H$-magnitude space, to determine what it can tell us about the $H$-distribution. 
To investigate the absolute magnitude-distribution, we implemented a four parameter $H$-magnitude distribution as parameterised by \citet{shankman13}: a divot, which for bright objects has an exponential increase with exponent $\alpha_b$, which extends to a transition $H$-magnitude at $H_t$ (which throughout this work will be in given in $r$-band, and was thus expected to occur near $H_t=8.5$), at which the number density has a drop by a contrast factor $c$, followed by another exponential function for faint objects with exponent $\alpha_f$; in case of $c=1$ the divot becomes the popular knee, and if in addition $\alpha_f=\alpha_b$, the knee becomes a single exponential. 
As we have fixed $\alpha_b$ to $0.95$ as found above, we now investigated the three-dimensional divot parameter space ($H_t$, $c$, $\alpha_f$) for Plutinos with $H_r<10$, thus comparing to the full set of 42 detected Plutinos. 
With the full set of data, we find that our observations favour (finds least rejectible) a moderately deep divot ($c=6$) at $H_t=8.4$ to a moderately steep slope ($\alpha_f=0.8$ for $H_r=8.4$ to $10.0$); see Fig. \ref{fig:contour-bestdivot} for a contour plot showing allowable ranges of the parameters. 
Our analysis rules out a single exponential (with $\alpha_f=\alpha_b=0.95$) at $>99\%$ confidence. 
While our analysis favours a moderately deep divot, a knee with $H_t<8.0$ to $\alpha_f\simeq0.2$-$0.4$ (as suggested for the hot TNOs by \cite{fraser14}) cannot be excluded. 
Fig. \ref{fig:cumul-bestdivot} shows the cumulative distribution of real and simulated detections from various models; to the eye, the divot clearly provides the best match to the observations. 
This is not surprising, given that our survey detected no Plutinos with $8.27<H_r<9.01$, yet found six Plutinos with $H_r\geq9.01$; these detections might even suggest that there is a shallow slope just after the drop, followed by a much steeper slope. 
Either way, a steep ($\alpha_f>0.6$) post-transition slopes cannot continue, because if such a slope is extrapolated indefinitely, the total mass of the trans-Neptunian region would diverge \citep{gladman01}. 
So, an $\alpha_f\simeq0.8$ post-divot slope is suitable over a limited $H$-magnitude range, but there there must be yet another transition at smaller sizes to a gentler slope. 
This could be consistent with the Plutino magnitude-distribution having the wavy form suggested by \citet{schlichting13}, although our data suggest the onset of the steepest portion of this wave pattern at larger sizes than the $D\approx10\unit{km}$ transition they estimate. 
Alternately, Fig. \ref{fig:contour-bestdivot} makes it clear that many $\alpha_f<0.6$ divot models are not rejectable, including the original ($H_t=8.5$, $c=6$, $\alpha=0.5$) proposal from \citet{shankman13}.

CFEPS predicted $13000^{+6000}_{-5000}$ Plutinos with $H_g<9.16$ \citep{gladman12}, corresponding to $H_r<8.66$, where uncertainties indicate the $95\%$ confidence range.
Our survey combined with CFEPS has 34 Plutinos with $H_r<8.66$, so we ran survey simulations of the combined surveys with our refined Plutino model cut off at $H_r=8.66$ until each simulation detected 34 synthetic objects. 
We counted the total number of generated objects required to provide these detections for each run in order to estimate the total Plutino population size and uncertainty.
With this boot-strap approach, we predict $9000\pm3000$ Plutinos with $H_r<8.66$, regardless of whether we use our favoured divot magnitude-distribution or the best knee distribution, showing a $\sim\frac{1}{\sqrt2}$ improvement in relative uncertainty as expected from roughly doubling the number of detections. 
This revised population estimate is in agreement with the CFEPS prediction, although this is not surprising due to the overlap of the sample. 
The addition of the new, deeper survey block described in this paper allows us to estimate the population of even fainter objects; we predict $37000^{+12000}_{-10000}$ Plutinos with $H_r<10.0$ assuming the divot distribution, and $35000^{+12000}_{-10000}$ if the knee is assumed. 
It may seem surprising that these two models produce similar population estimates, but this it because the knee happens sooner than the divot, so the divot has more $7.7<H_r<8.4$ objects than the knee, while the knee has more objects just past $H_r=8.4$, thus roughly equalising the total population estimate. If we could probe several magnitudes past the transition, the difference in population would become clearer. 

  \subsection{Two-bin test} \label{sec:twobin}

Above we found that the Plutinos must have a transition in their absolute magnitude-distribution, but both a divot and a knee provided non-rejectable models, when analysed using the Anderson-Darling method described in Sec. \ref{sec:stats}.
However, these two models should be measurably different in our data. 
We suspect that the reason the above method is failing to discriminate between the two models is that the AD test is biased to sensitivity towards the tails of distributions. 
The the main signature of a transition, as can be seen in Fig. \ref{fig:cumul-bestdivot}, should occur around the location of the transition. 
To test the models in an alternative, simple fashion, we constructed a ``two-bin test'' which focuses on the $H$-distribution just within two bins, in order to focus on the region around a transition. 
This test investigates the rejectability of a model $H$-magnitude distribution, given the number of detections in the $H$-mag bin immediately preceding and following a putative transition, ignoring all objects that are not in those two bins. 
This test is envisioned as a simple, intuitive attempt to get a qualitative sense of the nature of the data and which model to prefer. 
Focusing on only a small magnitude range should better reveal the location and nature of a transition.  

\begin{figure}[tbp]
\includegraphics[width=0.7\textwidth]{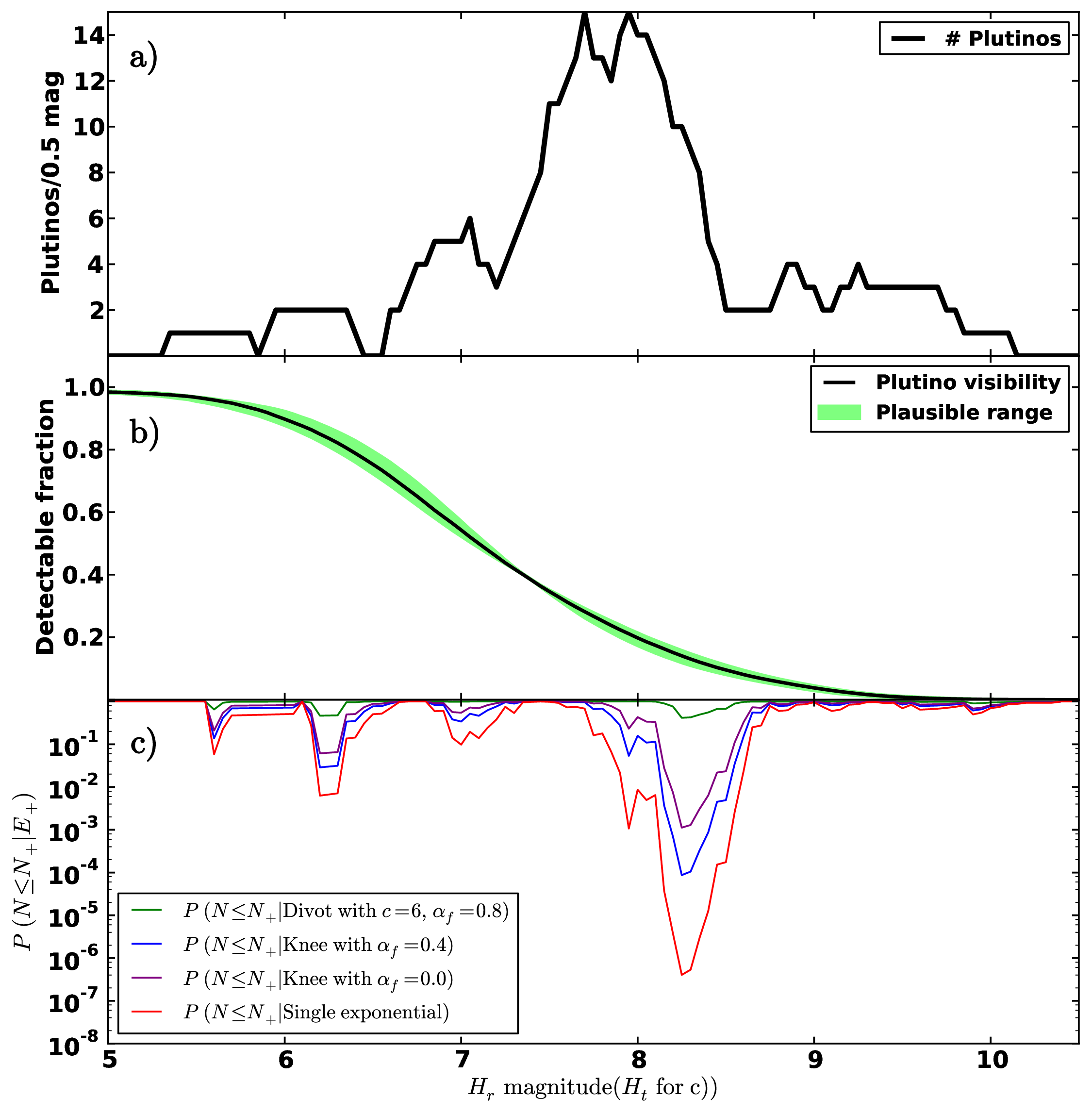}
\caption{\label{fig:twobin}
a) The number of Plutinos per 0.5 mag bin, at 0.05 mag intervals (to reveal binning-effects). \newline
b) The visibility of Plutinos to our combined surveys, assuming our favoured orbital-distribution (black curve), as a function of absolute magnitude.  
Note that the visibility only includes objects detected by the survey simulator, and thus already takes the detection efficiency and our limiting magnitude into account; the visibility is thus reliable even at small values. 
The green region shows the uncertainty in visibility generated from assuming different orbital-distributions within the error-range; the uncertainty is a small fraction of the visibility for $H<9$. \newline
c) Illustration of the two-bin test, showing $P(N\leq N_+|E_+)$, the probability of getting the observed number of detections or less in the next bin, given the expectation based on the previous bin. 
Low values thus shows that the model overpredicts the number of detections. 
}
\end{figure}

Fig. \ref{fig:twobin}a shows all of the 42 Plutinos from our combined surveys, as a number density per 0.5 magnitudes at a 0.05 magnitude intervals.
It is clear that there is an exponential rise in the number of detections from $H_r=5$ to roughly $8$, but then there is a precipitous drop in number of detections just past $H_r=8.0$. 
Could that drop be simply due to a sudden drop in sensitivity to fainter objects?  

To investigate this, we determined the ``visibility'' of the Plutinos as a function of $H$-magnitude.
We define the visibility as the fraction of objects with a given $H$-magnitude which, assuming an orbital-distribution, would be detected and tracked with a given survey; that is, the probability that, if the object is in the field of view, it is brigter than the survey characterisation limit. 
An object's visibility depends on its absolute magnitude, the object's orbit and the detection efficiency. 
A low-$H$ object will have a high visibility, as it is detectable even at apocentre; a moderage-$H$ object would have a smaller visibility, as it would only have a detectable apparent magnitude during part of its orbit; a very small (large-$H$) object will not even be detectable at pericentre and thus has a visibility of 0. 
The visibility of Plutinos to our combined surveys, given the our refined orbital model, was obtained by passing a flat $H$-mag distribution (ie. same number of objects at every $H$-magnitude) through the survey simulator, thus obtaining a fraction of objects detected as a function of $H$-mag. 
This fraction was then normalised as a visibility from 0 to 1 in order to only account for objects within the surveyed fields and is shown in Fig. \ref{fig:twobin}b. 
Unsurprisingly, the visibility \emph{cannot} have sharp feature for a population with a wide eccentricity-distributions like the Plutinos. 
Abrupt features in Fig. \ref{fig:twobin}a must thus be real variations in the $H$-mag distribution or statistical fluctuations. 
As we only used simulated objects with apparent magnitudes brighter than our $40\%$ survey characterisation limits to construct the visibility, we have confidence that the visibility is reliable even in the tail when it reaches small values. 
To confirm this, we constructed the visibility for a range of non-rejectable orbital models around our prefered one, to quantify the effect of slightly altering the orbital model. 
As can be seen in Fig. \ref{fig:twobin}b, none of the models can produce a sharp transition and all produce similar visibility curves. 

Having calculated the visibility, we perform the following algorithm, which we refer to as ``the two-bin test'', to study the evidence for a sharp transition at a candidate $H_t$: 
\begin{enumerate}
\item Bin objects into two $H$-mag bins; $B_-$ covering $[H_t-\Delta_H, H_t)$ and $B_+$ covering $[H_t, H_t+\Delta_H)$. 
The number of objects in these bins is $N_-$ and $N_+$, respectively.
\item Calculate $E_+$, the ``expectation value'' for the number of objects in $B_+$, given the known value of $N_-$, the assumed $H$-mag distribution model and the visibility-fraction in each bin as described above. 
\item Calculate the Poisson probability $P(N\leq N_+|E_+)$ of observing the detected number of objects or less in $B_+$, given the expectation value $E_+$. 
$$P(N\leq N_+|E_+) = \sum_{N=0}^{N_+} \frac{(E_+)^N\exp(-E_+)}{N!}$$
\item Shift $H_t$ by $\delta_H$ and start over. It is important that $\delta_H$ is significantly smaller than $\Delta_H$, in order to demonstrate that the choice of bin-centres and bin-widths are not crucial to any given result. 
\end{enumerate}
Low values of $P(N\leq N_+|E_+)$ reject the model at that $H_t$, as a low probability means that the model is overpredicting the number of detections (that is, the expected number is significantly greater than the observed number). 
For a single exponential function, a low value at a given $H_t$ means that the signle exponential function cannot explain the number of detections in the bin after $H_t$. 
When testing a divot or knee function, we used the divot and knee models derived in Sec. \ref{sec:bestmodel}, except for the determined $H_t$ of the transition. 
We let $H_t$ freely vary as the division point between the two bins, testing every value in the observed $H$-range; this way, the test gives an independent determination of where a transition occurs. 

The probabilities, given different models, as a function of the varying $H_t$ magnitude can be seen in Fig. \ref{fig:twobin}c. 
This plot appears to show a strong signature of a drastic transition around $H_t\sim8.3$, supporting a divot scenario (since we saw earlier that knees were prefered if $H_t <8.0$ while divots are prefered for $H_t >8.0$).
We also tested a knee to a flat slope (magenta curve in Fig. \ref{fig:twobin}c), such that $\alpha_b=0.95$ and $\alpha_f=0.0$. 
Even this knee presents evidence of overpredicting at $H_r~8.4$, showing that the transition must actually have a drop in numbers, that is, a divot. 
However, how strong is this evidence? 
In order to test the efficacy of this test, we repeated the same test on simulated detections (rather than our real detections), created by passing various models through our survey simulator. 

\begin{figure}[tbp]
\includegraphics[width=0.7\textwidth]{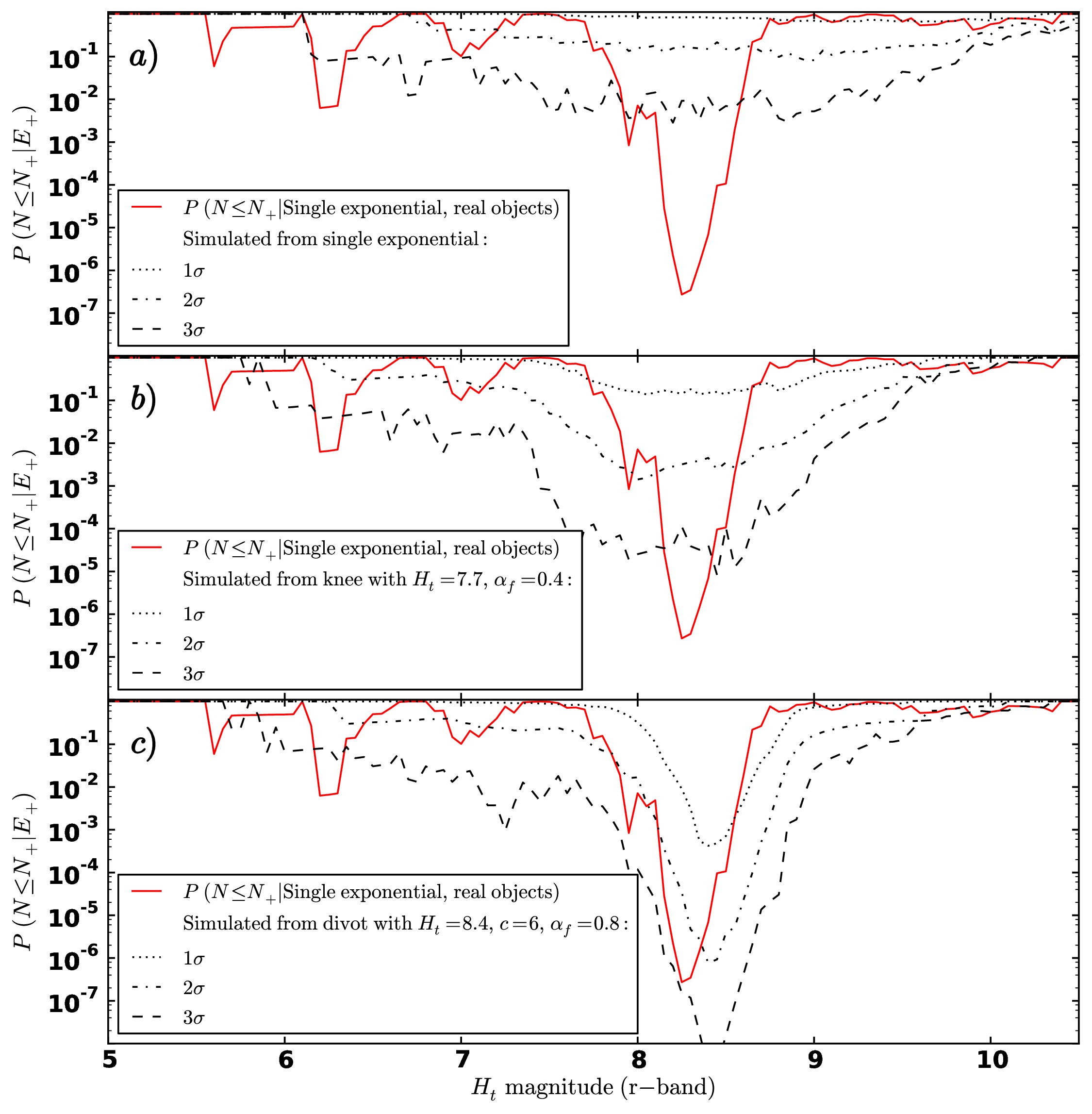}
\caption{\label{fig:efficacy}
Each plot shows the $68.2$ (dotted), $95.4$ (dash-dotted) and $99.7\%$ (dashed) bounds from testing a single exponential against 1000 sets of 42 simulated detections using the two-bin test. 
The simulated detections were generated from a) a single exponential magnitude-distribution with $\alpha_b=0.95$, b) a knee magnitude-distribution with $\alpha_b=0.95$, $H_t=7.7$, $\alpha_f=0.4$, c) a divot magnitude-distribution with $\alpha_b=0.95$, $H_t=8.4$, $c=6$, $\alpha_f=0.8$. 
In all three plots, the solid red line is the same as in Fig. \ref{fig:twobin}c, for reference. 
It is clear that in a) and b) $99.7\%$ of the simulated sets never generate signatures as strong as the one seen for the real data, making the hypothesis that ``a single exponential or knee magnitude-distribution could produce our real detections by chance'' rejectable at $>3\sigma$ level.
In c) however, the signal from our real data lies between the $95.4\%$ line and the $99.7\%$ line; while this is still in the tail, it is by far the best of the three, indicating that an intrinsic divot distribution is more likely to have produced our detections than a knee or a single exponential.
The fact that the centre of the dip seen for the simulated and real data is off by $0.1-0.2$ magnitudes might suggest that the intrinsic divot is in fact at $H_t=8.2-8.3$.
}
\end{figure}

Simulated detections were generated from a single exponential model with $\alpha=0.95$, to see whether detections drawn from such a model could cause signatures like those seen for the real detections in Fig. \ref{fig:twobin}c. 
This was done with 1000 simulated sets of detections of 42 Plutinos, to get a range of possible outcomes, and the 1, 2 and 3-$\sigma$ boundary was found. 
When testing a single exponential against simulated detections generated from a single exponential, the two-bin test should not give low probabilities; 1, 2 and 3-$\sigma$ outliers should have probabilities of $\sim32\%$, $\sim4.6\%$ and $0.3\%$. 
Fig. \ref{fig:efficacy}a confirms that this is indeed the case and that the simulated detections never cause the test to give probabilities as small as our real detections do. 
The  signature we see from the real detections thus cannot be caused by an underlying single exponential magnitude-distribution with $\alpha=0.95$, and some transition must be causing this signature, confirming our results from Sec. \ref{sec:bestmodel}. 

The process was repeated in a similar way, generating a thousand samples of 42 simulated detections using our favoured knee solution, with $\alpha_b=0.95$, $H_t=7.7$, $c=1$, $\alpha_f=0.4$. 
The two-bin test was performed on these simulated detections, in order to see whether a knee in the intrinsic distribution can explain the signature seen in the real detections. 
Here we only present testing a single-exponential against the simulated detections, as testing the other models do not add significant information and would only clutter the figures. 
We are thus always considering whether the simulated detections could explain the signature seen when testing a single exponential against the real detections.
Although there is improvement when the simulated detections are drawn from a knee, as seen in Fig. \ref{fig:efficacy}b, even this magnitude-distribution cannot quite explain the signature that the real data gives us. 
The transition must be more drastic than the knee.

Lastly, we used our favoured divot magnitude-distribution, with $\alpha_b=0.95$, $H_t=8.4$, $c=6$, $\alpha_f=0.8$, to generate simulated detections and subsequently test a single exponential against the simulated detections using the two-bin test. 
With the simulated detections from the divot, the 1, 2 and 3-sigma curves from the two-bin test (Fig. \ref{fig:efficacy}c) has a very deep dip at the transition magnitude, similar to that seen for the real detections. 
While the agreement is not perfect, the signature of the real and simulated detections are far more similar than any of the above, suggesting that a divot is a better model for the observed signature. 
We take this as reasonable evidence that the Plutino magnitude-distribution has a divot. 

\section{Neptunian Trojans analysis}\label{sec:trojans}

Our survey discovered two Neptunian Trojans, mah01=2012 UW$_{177}$ and mah02=2012 UV$_{177}$ (see Table \ref{tab:objects}). 
CFEPS \citep{petit11} discovered one, 2004 KV$_{18}$, which was recognised as a temporary Trojan by \cite{hornerlykawka12}.
While some constraints on the absolute magnitude-distribution based on three objects might be possible, there exists the added complication that there are two populations of Neptunian Trojans: the long term stable Trojans (stable on 4 Gyr time scales) and the short-term captured Trojans (unstable on kyr-Myr time scales). 
There are currently 11 known Neptunian Trojans\footnote{List of Neptunian Trojans in the MPC database as of 2013 Nov 11: \url{<http://web.archive.org/web/20131112211347/http://www.minorplanetcenter.org/iau/lists/NeptuneTrojans.html>}}, of which two are certain to be temporary (\citet{hornerlykawka12},supplementary material of \citet{alexandersen13b}, this work) and two have uncertain stability \citep{brasser04,hornerlykawka10a,horner12,guan12}. 
These 11 were discovered in several different surveys, some of which do not have publicly available characteristics, making it impossible to analyse the combined set reliably. 
The following analysis therefore only used detections from our new survey and CFEPS. 
Assuming the dynamically hot TNO populations share a common origin and thus have the same size-distribution \citep{shankman13} the distinction between stable and temporary Neptunian Trojans is simply the difference in the eccentricity and inclination-distributions. 
We assumed that the temporary co-orbitals and stable trojans share an $H$-magnitude distribution with the hot TNOs; we have therefore applied absolute magnitude-distributions of same form as derived for the Plutinos in order to produce population estimates. 
We have, however, analysed the two components (stable and temporary) separately. 

  \subsection{Stable Neptunian Trojans}

Our survey found one stable Neptunian Trojan, 2012 UV$_{177}$, with $H_r=8.93$. 
Via 4 Gyr numerical integrations, using the range of orbits established by the orbital classification algorithm of \citet{gladman08}, we find this to be a secure stable 1:1 Trojan resonator, with a libration amplitude around the L4 point of $13\degree$. 
The fact that 2012 UV$_{177}$ has an inclination of $21\degree$ and the fact that we discovered no stable Neptunian Trojans in our low-lat block continues to strengthen the assertion that the stable Neptunian Trojans have a very dynamically hot inclination-distribution \citep{sheppardtrujillo06,sheppardtrujillo10a}, with recent indications that the standard $15\degree$ width of other hot populations is still plausible, although at the low end of the allowable range \citep{parker14}.

\begin{figure}[tbp]
\includegraphics[width=0.8\textwidth]{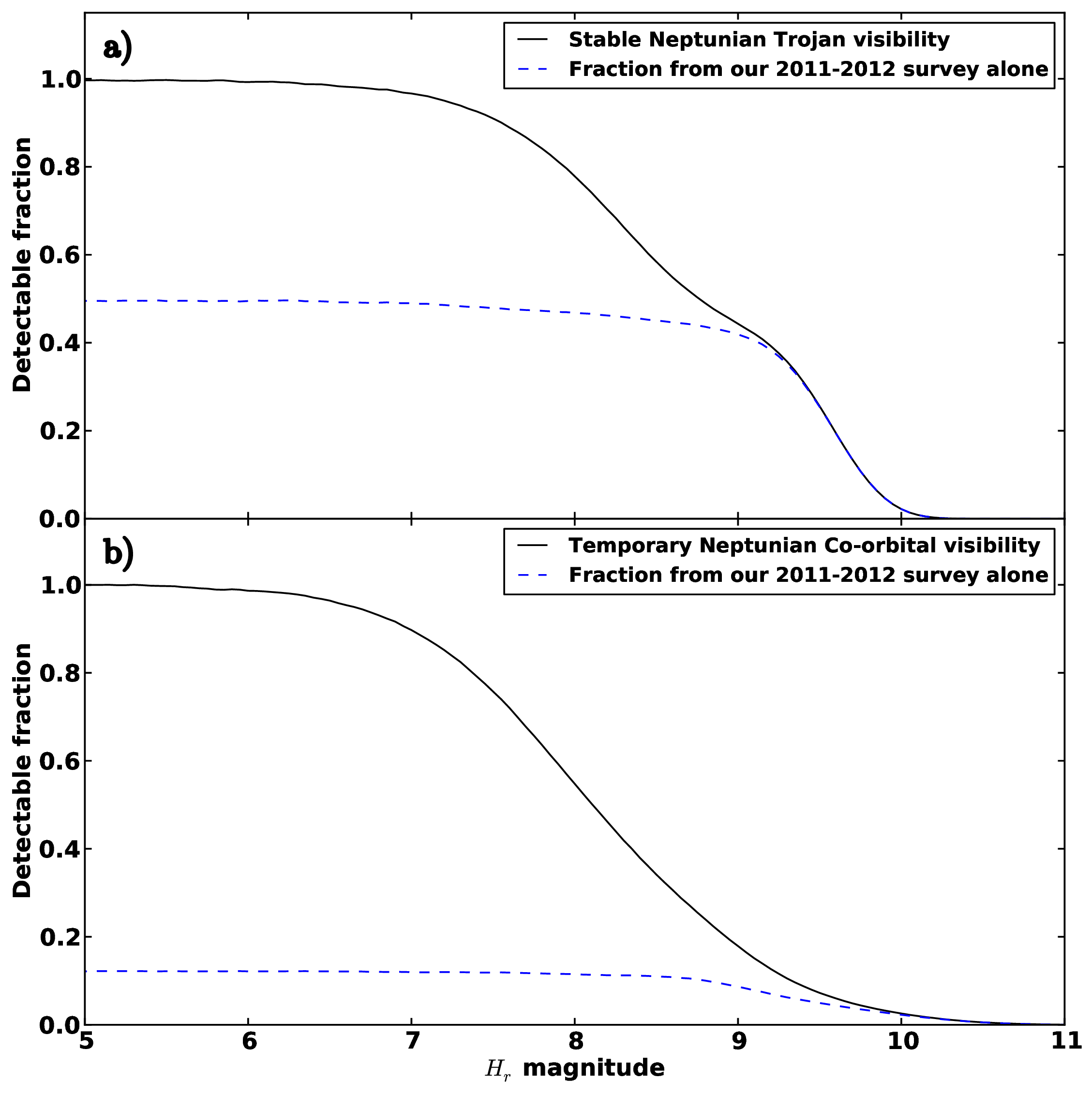}
\caption{\label{fig:visibility_NT+NC}
a) The visibility of stable Neptunian Trojans (black curve) to the combined surveys, assuming the \citet{parker14} intrinsic orbital-distribution, as a function of absolute magnitude. 
The blue dashed curve shows the visibility contributed by our new survey alone, which, due to being pointed very close to the centre of the Trojan cloud, was as sensitive to bright Trojans as the entirity of CFEPS was, despite the latter's larger areal coverage. 
Our survey's greater depth is also evident. \newline
b) The visibility of temporary Neptunian co-orbitals (black curve) to the combined surveys, assuming the \citet{alexandersen13b} orbital-distribution for temporary Neptunian co-orbitals, as a function of absolute magnitude. 
Our new survey contributes a much smaller fractions here (blue dashed curve) compared to the larger coverage of CFEPS, because the temporary co-orbitals have larger libration amplitudes. 
Our new survey still probes fainter objects, supplying the majority of the sensitivity in the tail beginning at $H_r\approx9.0$.
}
\end{figure}

Using the favoured orbital-distribution derived by \citet{parker14} for the stable Neptunian Trojans, we generated population estimates (Table \ref{tab:popest}) and upper limits for various $H_r$-cuts using two different $H$-distributions (the divot and knee favoured by our Plutino analysis).
The visibility of Neptunian Trojans, given this orbital-distribution, can be seen in Fig. \ref{fig:visibility_NT+NC}a, showing that our new survey was as sensitive to Neptunian Trojans as CFEPS, despite our smaller sky-coverage, because our fields were chosen to optimise Neptunian Trojan detection. 

\citet{sheppardtrujillo10b} estimated about 400 Neptunian Trojans (stable + temporary) with $D>100\unit{km}$, which corresponds to $H_r<8.6$ given their assumptions and appropriate colour conversion.
CFEPS \citep{gladman12} put an upper limit of 300 ($95\%$ confidence) stable Neptunian Trojans with $H_g<9.16$ (chosen to correspond with $D>100\unit{km}$), which (assuming colours similar to the Plutinos) correspond to $H_r<8.66$. 
Our $95\%$ confidence upper limit, $<250$ objects with $H_r<8.66$, is even lower than previous estimates and disagrees with the \citet{sheppardtrujillo10b} prediction of 400. 
However, part of this difference is due the fact that only seven of the nine objects used in that work have in fact turned out to be Trojans. 

Our cumulative estimated populations for $H_r<9.1$ and $10.0$, $80^{+300}_{-70}$ and $150^{+600}_{-140}$ respectively, are also lower than the \cite{sheppardtrujillo10b} prediction, indicating that the stable Neptunian Trojans are not nearly as numerous as the initial estimates, and thus less numerous than the main asteroid belt, which has over 700 asteroids with $H_r<10.0$ \citep{jedicke02}. 

  \subsection{Temporary Neptunian co-orbitals}

Temporary capture from scattering/Centaur orbits into 1:1 resonance with Neptune has been found, perhaps surprisingly, to be frequent \citep{alexandersen13b}. 
In addition to tadpole libration (Trojans), there are other types of temporary 1:1 mean motion resonance orbits, namely horseshoe and quasi-satellite orbits. 
Here we have grouped them all together as temporary Neptunian co-orbitals. 

Our survey has found one temporary Neptunian co-orbital, 2012 UW$_{177}$. 
With $e=0.259$ and $i=53.89\degree$, 2012 UW$_{177}$ has the most extreme orbit of the Neptunian co-orbitals discovered to date. 
At $H_r=10.65$ it is also the intrinsically faintest (so likely smallest); assuming $5\%$ albedo, its diameter is $\approx40\unit{km}$.
Applying the classification algorithm of \citet{gladman08} shows that although this object is currently librating around Neptune's leading L4 Lagrange point (executing tadpole motion ranging from 35--75 degrees ahead of Neptune), in less than 1.5 Myr the long-term secular cycle of its eccentricity causes it to dynamically couple with Uranus.
Because the 1:1 resonance with Neptune provides no protection mechanism against Uranus encounters, the object is then perturbed out of the resonance to scattering behaviour.
Thus it appears plausible that as recently as 200,000 years ago (when our integrations show the eccentricity was again sufficiently high for Uranus coupling) the object was a scattering object (or Centaur if $a<30\unit{AU}$) that had a Uranus encounter which perturbed the orbital semi-major axis to that near Neptune, allowing the 1:1 resonance dynamics to begin. 
The resonance dynamics protected the object from Neptune encounters, while a secular cycle in the pericentric distance $q$ protected the object from Uranus for a Myr ($q$ is currently rising through 22.3 AU to a maximum near 28 AU about 400 kyr from now).
However, our integrations always show that all three clones will recouple to Uranus during the next eccentricity maximum about 1.5 Myr from now and leave the tadpole and co-orbital region.
The past and future behaviour of this temporarily captured object is the same as the dynamics discussed in \citet{alexandersen13b} with the only noteworthy feature being the exceedingly large orbital inclination.
Due to the large planetary encounter speeds produced by the large $i$, the perturbations by both Uranus and Neptune are weak, resulting in small $(a,q,i)$ changes in any single encounter; this implies that this small body was likely originally delivered to $a\simeq$30~AU with $i>50\degree$ already. 
The source of these large-inclination TNOs remains unclear \citep{gladman09b,brasser12,volkmalhotra13}.

For the temporary Neptunian co-orbitals, we obtained population estimates using the orbital-distribution of co-orbitals captured from the scattering population calculated in \citet{alexandersen13b}. 
A visibility plot for such temporary Neptunian co-orbitals can be seen in Fig. \ref{fig:visibility_NT+NC}b. 
Our survey contributes a smaller fraction to the visibility of these objects compared to CFEPS, because many have large libration amplitudes and are thus detectable in a larger ecliptic longitude range. 
Our survey does extend the tail of the $H$-magnitude visibility, which resulted in the detection of 2012 UW$_{177}$ in the very tail.
CFEPS also discovered one temporary Neptunian co-orbital, 2004 KV$_{18}$, which has $H_r\sim8.2$ and is currently on an L5 tadpole orbit that is unstable on $\sim0.2\unit{Myr}$ timescale \citep{hornerlykawka12}, so our population estimates are based on two objects (Table \ref{tab:popest}). 
Surprisingly, the number of stable Neptunian Trojans and temporary co-orbitals appear to be very similar.
The dynamical lifetimes of the temporary co-orbitals are less than $0.1\%$ of the age of the Solar System, so given similar population numbers, it is impossible for the temporary Neptunian co-orbitals to all be previously stable Neptunian Trojans that just happen to be leaking out of the stability zone now; they must be captured from a different reservoir. 
The very large inclination of 2012 UW$_{177}$ should therefore not be taken as an indication of the inclination-distribution of the stable Neptunian Trojans, which with a width of $\sim15-20\degree$ should have few if any of such high inclination objects. 
In the model where they are captured from the scattering population, the temporary Neptunian co-orbitals could thus plausibly have an even hotter inclination-distribution than the stable Neptunian Trojans. 

With just two objects, the magnitude-distribution of the temporary Neptunian co-orbitals cannot be statistically constrained, as a knee or divot, or even a steep single exponential with $\alpha=0.95$, cannot be rejected. 
However, a knee appears to be the worst of the three, presumably because of the shallow slope for faint objects. 
It is noteworthy that a significant fraction of both our temporary Neptunian Trojan and Plutino detections were very close to our visibility limit; six of our Plutinos have $H_r>9.0$, where their visibility is less than $10\%$, implying that there is an abundance of these very faint objects. 
It is possible that, in the divot and just after, we are seeing some evidence of a transition to the very steep section of the size-distribution predicted by \citet{schlichting13}.

\section{Uranian Trojans - population estimate} \label{sec:uranian}

With only one temporary Uranian co-orbital in our sample, mal01=2011 QD$_{99}$ (there was none in CFEPS), we simply make an order of magnitude population estimate of temporary Uranian co-orbitals (Table \ref{tab:popest}), again based on their orbital-distribution computed in \citet{alexandersen13b}. 
While the large uncertainty ranges overlap, the most likely estimates suggest that there are fewer temporary Uranian co-orbitals than Neptunian ones, in agreement with the theoretical prediction \citep{alexandersen13b} that, in steady state, the temporary Neptunian co-orbitals should outnumber the temporary Uranian co-orbitals by a factor of 7. 

\section{3:1 and 4:1 resonators - population estimates} \label{sec:fourtinos}

Our survey discovered three objects in the 3:1 outer resonance with Neptune, and CFEPS discovered one. 
We used the simply CFEPS L7 model of the 3:1 resonance to determine a population estimate. 
However, in order to not make unfounded predictions, we constrain the eccentricity distribution (which is otherwise flat) such that the pericentre of synthetic objects is greater than $q=31.2\unit{AU}$ (pericentre distance of our most eccentric object, with eccentricity $e=0.50$) and so that the smallest synthetic eccentricity was equal to the smallest observed eccentricity, $=0.40$. 
With this model, we determine that there are $3000^{+4000}_{-2000}$ objects in the 3:1 resonance with $H_r<8.66$, $e>0.4$ and $q>31.2\unit{AU}$.

Additionally, our survey discovered what we believe is first object to be securely determined to be in the 4:1 resonance with Neptune, 2011 UP$_{411}$. 
\citet{emelyanenkokiseleva08} estimates the probability that 2003 LA$_7$ and 2000 PH$_{30}$ are in the 4:1 resonance to be $11\%$ and $4\%$, respectively, while finding $<1\%$ probability that 2005 TB$_{190}$ and 2000 PF$_{30}$ are resonating. 
Based on zero detections in CFEPS, \citet{lawler13} predicted a $95\%$ confidence upper limit for the 4:1 population of $16,000$ with $H_g<9.16$ (or $H_r<8.66$ assuming $g-r=0.5$). 
We took the model used above for the 3:1 resonance and adjusted it appropriately for the 4:1 resonance, with semi-major axis $a=75.8$. 
The lower bound on the peri-centre distance was kept at $31.2\unit{AU}$ (as we are of course sensitive to closeby objects) and the bound on the eccentricity was changed be the eccentricity of 2011UP$_{411}$, $e=0.49$. 
Using this model determine that the 4:1 resonance has $2500^{+10000}_{-2400}$ objects with $H_r<8.66$, $e>0.49$ and $q>31.2\unit{AU}$. 

This shows that even just the high-eccentricity part of the 3:1 and 4:1 resonances are very populous (an order of magnitude more populous than the Uranian and Neptunian Trojans and temporary co-orbitals). 
It is possible that these populations have an as-yet undetectable low-eccentricity component, making them potentially as numerous as the Plutinos (Table \ref{tab:popest}). 
Recent results \citep{pike15} indicate that the 5:1 resonance is far more populous than the Plutinos. 
The high population numbers of these distant resonances are intriguing, as there is no known mechanism for populating these distant, high-order resonances \citep{chiang03,hahnmalhotra05,levison08,gladman12}. 

\section{Conclusion}

We have presented a detailed description of a carefully characterised survey carried out in 2011-2013. 
This includes a full list of detected objects and a detailed description of our survey characterisation, allowing our survey to be utilised by further theoretical work by others. 

Our survey detected 87 TNOs, 77 of which were brighter than our survey's characterisation limit. 
All but two objects from the characterised sample were tracked to two-opposition arcs. 
Our observing cadence and improved astrometric techniques allowed secure classification of most objects from these two-opposition arcs. 
Of particular interest was one temporary Uranian Trojan, two Neptunian Trojans (one stable and one temporary), 18 Plutinos, three 3:1 resonators and a 4:1 resonator. 
Population estimates and upper limits for these populations were presented, as well as a detailed analysis of the Plutino absolute magnitude-distribution. 

Using 24 Plutinos from CFEPS and 18 Plutinos from our new survey, we found that the CFEPS orbital-distribution model for Plutinos is acceptable and make only minor adjustments to the model, primarily in lowering the mean inclination. 
The exponential absolute magnitude-distribution cannot continue past $H_r\sim8.4$, as the expected number of faint detections would then vastly outnumber the observed number. 
Knee and divot distributions were investigated to reveal ranges of allowable parameters. 
We found that our data favours a divot at $H_r=8.4$ with pre-divot slope of $0.95$, post-divot slope of $0.8$, and divot contrast of $6$, while knees at $H_r<8.0$ to a shallower slope cannot be ruled out. 
However a secondary test of the data again favours the divot scenario. 

While we cannot set constraints on the magnitude-distribution of the Neptunian and Uranian co-orbitals, stable or temporary, we find that the population of stable and temporary Neptunian co-orbitals are comparable in number, indicating that the stable Neptunian Trojans cannot be the source population of the temporary Neptunian co-orbitals. 

\section{Acknowledgements}
\acknowledgments

We would like to thank R. Pike and S. Lawler for valuable feedback on this manuscript. 
We would also like to thank the CFHT queue service observers for their help and effort in making this project a success. 

M. Alexandersen and B. Gladman were supported by the National Sciences and Engineering Research Council of Canada. 

This work is based on observations obtained with MegaPrime/MegaCam, a joint project of CFHT and CEA/IRFU, at the Canada-France-Hawaii Telescope (CFHT) which is operated by the National Research Council of Canada, the Institut National des Science de l'Univers of the Centre National de la Recherche Scientifique of France, and the University of Hawaii.

This work uses observations obtained at the Gemini Observatory, which is operated by the Association of Universities for Research in Astronomy, Inc., under a cooperative agreement with the NSF on behalf of the Gemini partnership: the National Science Foundation (United States), the National Research Council (Canada), CONICYT (Chile), the Australian Research Council (Australia), Minist\'{e}rio da Ci\^{e}ncia, Tecnologia e Inova\c{c}\~{a}o (Brazil) and Ministerio de Ciencia, Tecnolog\'{i}a e Innovaci\'{o}n Productiva (Argentina).

This publication makes use of data products from the Two Micron All Sky Survey, which is a joint project of the University of Massachusetts and the Infrared Processing and Analysis Center/California Institute of Technology, funded by the National Aeronautics and Space Administration and the National Science Foundation.

{\it Facilities:} \facility{CFHT (MegaCam)}, \facility{Gemini:Gillett (GMOS)}.

\bibliographystyle{apj}
\bibliography{a14m}

\clearpage

\clearpage

\begin{deluxetable}{llllll}
\tabletypesize{\scriptsize}
\tablecaption{Field coverage of our survey. 
The width and height describe a rectangle centred on the given RA and Dec. 
JD is the date of the discovery observations.
The fill factor is the fraction of the area inside the rectangle that was covered (less than 1 due to chip gaps and separation of fields).  
\label{tab:pointings}}
\tablehead{
\colhead{Width ($\degr$)} & \colhead{Height ($\degr$)}   & \colhead{RA} & \colhead{Dec} & \colhead{JD} & \colhead{Fill factor ($\%$)}
}
\startdata
\multicolumn{6}{c}{\textbf{Low-latitude block}}\\
 4.930 & 0.995 & 01:51:08.00 & +13:24:48.0 & 2455858.963195 & 90.9\\
 4.914 & 0.995 & 01:51:08.00 & +14:23:36.0 & 2455858.963195 & 91.2\\
 4.896 & 0.995 & 01:51:08.00 & +15:22:24.0 & 2455860.843750 & 91.5\\
 4.878 & 0.995 & 01:51:08.00 & +16:21:12.0 & 2455860.843750 & 91.6\\
\multicolumn{6}{c}{\textbf{High-latitude block}}\\               
 1.993 & 2.984 & 01:41:01.70 & +28:10:00.0 & 2456220.904167 & 90.7\\
 1.993 & 2.984 & 01:49:58.30 & +28:10:00.0 & 2456221.947917 & 90.7\\
\enddata
\end{deluxetable}

\clearpage

\begin{deluxetable}{lrrrrrrrrrrrc}
\tabletypesize{\scriptsize}
\tablecaption{Efficiency function parameters and characterisation limits of our survey blocks in the searched rate ranges. 
The efficiency function is parameterized in Eq. \ref{eq:eff}. 
The limit is the magnitude at which the efficiency hits $40\%$ of the maximum value. 
The functions are plotted in Fig. \ref{fig:efficiency}.
\label{tab:limits}}
\tablehead{
\colhead{Rate range (``/hr)} & \colhead{$f_{21}$}   & \colhead{k} & \colhead{$m_e$} & \colhead{$w$} & \colhead{Limit}
}
\startdata
\multicolumn{5}{c}{\textbf{Low-latitude block}}\\
0.50-2.06  & 0.96 & 0.0148 & 24.62 & 0.125 & 24.62 \\
2.06-3.72  & 0.91 & 0.0118 & 24.60 & 0.133 & 24.61 \\
3.72-5.38  & 0.90 & 0.0106 & 24.55 & 0.142 & 24.57 \\
5.38-10.00  & 0.88 & 0.0136 & 24.48 & 0.130 & 24.49 \\
\multicolumn{5}{c}{\textbf{High-latitude block}}\\
0.50-2.06  & 0.92 & 0.0194 & 24.64 & 0.118 & 24.60 \\
2.06-6.00  & 0.90 & 0.0133 & 24.59 & 0.115 & 24.60 \\
6.00-10.36 & 0.88 & 0.0139 & 24.49 & 0.124 & 24.49 \\
\enddata
\end{deluxetable}

\clearpage

Table \ref{tab:objects}.
List of all objects detected in our survey.
$ID$ is the internal designation, ``ma'' denoting the survey while ``l'' and ``h'' distinguish whether an object was discovered in the low-lat or high-lat block, respectively. 
$MPC$ is the Minor Planet Center designation.
The reduced Julian date and heliocentric distance at discovery are $Disc$ and $r$, respectively.
$N$ and $t$ are the total number of astrometric measurements and the total arc in years, respectively.
The bariocentric orbital elements, semi-major axis ($a$), eccentricity ($e$), inclination ($i$), longitude of ascending node ($\Omega$), argument of pericentre ($\omega$) and mean anomaly ($M$), are all given at the reduced Julian date epoch, $Epoch$.
The discovery magnitude $m_r$ is the average $r$-band apparent magnitude as measured by the moving object pipeline in the discovery triplet.
$H_r$ is the absolute magnitude calculated from $m_r$ using appropriate phase-angle corrections.
S denotes whether or not the classification is secure (S), insecure (I) or whether the object was never classified (N).
Note that three objects were previously discovered objects that happened to be in our field; all information in this work, appart from the MPC designation, ignores the fact that these objects were previously discovered, and the information in this table is therefore derived solely from this work.
The MPC Minor Planet electronic Circular containing the discovery astrometry next to each class header. 
The new astrometry for the three previously known TNOs is in \citet{tomatic14a}.
\setlength{\tabcolsep}{0.25em}
\begin{deluxetable}{l|l|c|c|r|c|c|c|c|c|c|c|c|r|r|c}
\rotate
\tabletypesize{\tiny}
\tablecaption{\label{tab:objects}
List of all objects detected in our survey.
}
\tablehead{
\colhead{$ID$} & \colhead{$MPC$} & \colhead{$Disc$} & \colhead{$r$ (AU)} & \colhead{$N$} & \colhead{$t$ (yr)} & \colhead{$Epoch$} & \colhead{$a$ (AU)} & \colhead{$e$} & \colhead{$i$ ($^{\circ}$)} & \colhead{$\Omega$ ($^{\circ}$)} & \colhead{$\omega$ ($^{\circ}$)} & \colhead{$M$ ($^{\circ}$)} & \colhead{$m_r$} & \colhead{$H_r$} & \colhead{S}
}
\startdata
\multicolumn{16}{c}{\textbf{Temporary Uranian Trojan} \citep{alexandersen13a}}\\
mal01     & 2011 QF$_{99}$  & 55858.9 & $20.296\pm0.000$ & 33 & 1.4722 & 55803.0 & $19.092\pm 0.001$ & $0.17687\pm0.00009$ & $10.810\pm 0.000$ & $222.499\pm  0.000$ & $287.47 \pm   0.02 $  & $258.64 \pm 0.02 $ & $22.6\pm0.1$ &  9.57 & S \\
(Winnie)  &                 &	      &		         &    &	       &	 &		     &		           &		       &		     &		             &		          &	         &       &   \\
\multicolumn{16}{c}{\textbf{Stable Neptunian Trojan} \citep{alexandersen14o}}\\
mah02     & 2012 UV$_{177}$ & 56220.9 & $29.578\pm0.001$ & 30 & 1.4614 & 55805.9 & $30.024\pm 0.004$ & $0.0723 \pm0.0009 $ & $20.833\pm 0.000$ & $265.668\pm  0.002$ & $204.28 \pm   0.10 $  & $285.94 \pm 0.02 $ & $23.7\pm0.1$ &  8.93 & S \\
\multicolumn{16}{c}{\textbf{Temporary Neptunian Trojan} \citep{tomatic14b}}\\
mah01     & 2012 UW$_{177}$ & 56220.9 & $22.432\pm0.001$ & 17 & 1.0429 & 56158.0 & $30.072\pm 0.003$ & $0.2591 \pm0.0002 $ & $53.886\pm 0.001$ & $20.010 \pm  0.001$ & $34.4   \pm   0.2  $  & $351.967\pm 0.004$ & $24.2\pm0.1$ & 10.65 & S \\
\multicolumn{16}{c}{\textbf{Resonant 4:3} \citep{alexandersen14n}}\\
mal09     & 2011 UZ$_{412}$ & 55858.9 & $31.606\pm0.001$ & 24 & 1.4422 & 55803.0 & $36.433\pm 0.003$ & $0.13295\pm0.00012$ & $6.812 \pm 0.001$ & $9.788  \pm  0.003$ & $27.4   \pm   0.7  $  & $355.932\pm 0.001$ & $24.0\pm0.1$ &  9.06 & S \\
mal10     & 2011 UA$_{413}$ & 55860.8 & $31.794\pm0.003$ & 19 & 1.2866 & 55859.8 & $36.528\pm 0.009$ & $0.1329 \pm0.0010 $ & $5.437 \pm 0.000$ & $292.353\pm  0.013$ & $83     \pm   2    $  & $11.076 \pm 0.005$ & $23.3\pm0.1$ &  8.29 & S \\
\multicolumn{16}{c}{\textbf{Resonant 3:2} \citep{alexandersen14a}}\\
mal45     & 2011 UD$_{411}$ & 55860.8 & $41.480\pm0.001$ & 28 & 2.2668 & 55804.0 & $39.21 \pm 0.03 $ & $0.2884 \pm0.0011 $ & $15.700\pm 0.001$ & $19.081 \pm  0.001$ & $131.88 \pm   0.11 $  & $274.68 \pm 0.09 $ & $23.1\pm0.1$ &  6.91 & S \\
mal04     & 2011 UA$_{411}$ & 55860.8 & $29.828\pm0.001$ & 29 & 2.1003 & 55804.0 & $39.223\pm 0.004$ & $0.24459\pm0.00009$ & $8.645 \pm 0.000$ & $237.962\pm  0.002$ & $167.51 \pm   0.09 $  & $351.158\pm 0.001$ & $23.7\pm0.1$ &  9.01 & S \\
mal02     & 2011 UC$_{411}$ & 55860.8 & $28.724\pm0.000$ & 43 & 2.0894 & 55804.0 & $39.270\pm 0.003$ & $0.27179\pm0.00009$ & $4.652 \pm 0.000$ & $289.175\pm  0.005$ & $88.35  \pm   0.06 $  & $6.447  \pm 0.001$ & $22.3\pm0.1$ &  7.74 & S \\
mal03     & 2011 UU$_{410}$ & 55858.9 & $28.997\pm0.001$ & 19 & 2.0893 & 55803.0 & $39.293\pm 0.004$ & $0.26294\pm0.00007$ & $12.623\pm 0.001$ & $19.330 \pm  0.001$ & $15.39  \pm   0.08 $  & $356.454\pm 0.001$ & $24.4\pm0.1$ &  9.85 & S \\
mah03     & 2012 UG$_{177}$ & 56220.9 & $30.482\pm0.002$ & 18 & 1.2344 & 55889.9 & $39.304\pm 0.010$ & $0.2248 \pm0.0003 $ & $21.876\pm 0.001$ & $261.872\pm  0.002$ & $130.6  \pm   0.7  $  & $357.684\pm 0.001$ & $24.4\pm0.1$ &  9.48 & S \\
mal30     & 2004 VT$_{75}$  & 55858.9 & $37.958\pm0.001$ & 31 & 2.2717 & 55803.0 & $39.31 \pm 0.02 $ & $0.2108 \pm0.0010 $ & $12.842\pm 0.001$ & $26.610 \pm  0.000$ & $272.04 \pm   0.03 $  & $68.66  \pm 0.05 $ & $21.9\pm0.1$ &  6.11 & S \\
mal16     & 2011 UY$_{410}$ & 55860.8 & $34.137\pm0.002$ & 24 & 1.4313 & 55804.0 & $39.312\pm 0.008$ & $0.1364 \pm0.0007 $ & $8.590 \pm 0.001$ & $12.404 \pm  0.002$ & $2.9    \pm   0.8  $  & $13.169 \pm 0.005$ & $24.4\pm0.1$ &  9.09 & S \\
mah15     & 2012 UH$_{177}$ & 56221.9 & $45.464\pm0.002$ & 27 & 1.4559 & 55806.0 & $39.39 \pm 0.02 $ & $0.1907 \pm0.0012 $ & $23.749\pm 0.001$ & $255.149\pm  0.003$ & $346.2  \pm   0.6  $  & $137.39 \pm 0.10 $ & $24.2\pm0.2$ &  7.57 & S \\
mal07     & 2011 UQ$_{410}$ & 55858.9 & $30.974\pm0.001$ & 26 & 2.0071 & 55803.0 & $39.400\pm 0.008$ & $0.2636 \pm0.0003 $ & $3.246 \pm 0.001$ & $10.821 \pm  0.004$ & $335.40 \pm   0.06 $  & $26.946 \pm 0.008$ & $24.2\pm0.1$ &  9.36 & S \\
mal21     & 2011 UX$_{410}$ & 55860.8 & $36.258\pm0.001$ & 28 & 2.1738 & 55804.0 & $39.40 \pm 0.02 $ & $0.3279 \pm0.0007 $ & $17.479\pm 0.001$ & $223.962\pm  0.000$ & $263.515\pm   0.011$  & $302.35 \pm 0.04 $ & $23.7\pm0.1$ &  8.14 & S \\
mal06     & 2011 UR$_{410}$ & 55858.9 & $30.875\pm0.001$ & 21 & 2.2693 & 55803.0 & $39.412\pm 0.009$ & $0.2685 \pm0.0004 $ & $17.320\pm 0.001$ & $27.050 \pm  0.000$ & $50.37  \pm   0.09 $  & $332.881\pm 0.009$ & $24.4\pm0.1$ &  9.59 & S \\
mal05     & 2011 UV$_{410}$ & 55858.9 & $29.862\pm0.001$ & 26 & 2.1766 & 55803.0 & $39.421\pm 0.007$ & $0.2591 \pm0.0003 $ & $4.308 \pm 0.000$ & $332.153\pm  0.006$ & $30.14  \pm   0.09 $  & $15.396 \pm 0.004$ & $22.5\pm0.1$ &  7.77 & S \\
mal17     & 2011 US$_{410}$ & 55858.9 & $35.150\pm0.001$ & 27 & 2.1028 & 55803.0 & $39.510\pm 0.013$ & $0.2554 \pm0.0005 $ & $11.876\pm 0.001$ & $20.578 \pm  0.001$ & $290.41 \pm   0.02 $  & $51.18  \pm 0.03 $ & $23.7\pm0.1$ &  8.27 & S \\
mal35     & 2011 UT$_{410}$ & 55858.9 & $39.355\pm0.001$ & 24 & 2.2420 & 55803.0 & $39.52 \pm 0.02 $ & $0.1787 \pm0.0011 $ & $13.498\pm 0.001$ & $19.796 \pm  0.001$ & $109.09 \pm   0.04 $  & $281.59 \pm 0.05 $ & $21.5\pm0.1$ &  5.58 & S \\
mal31     & 2011 UZ$_{410}$ & 55860.8 & $38.402\pm0.001$ & 26 & 2.3184 & 55804.0 & $39.521\pm 0.013$ & $0.1245 \pm0.0012 $ & $3.833 \pm 0.000$ & $298.461\pm  0.010$ & $8.44   \pm   0.04 $  & $69.87  \pm 0.04 $ & $24.0\pm0.1$ &  8.17 & S \\
mal28     & 2011 UW$_{410}$ & 55860.8 & $37.600\pm0.001$ & 30 & 2.1631 & 55804.0 & $39.578\pm 0.014$ & $0.2175 \pm0.0007 $ & $10.752\pm 0.001$ & $17.914 \pm  0.001$ & $285.455\pm   0.014$  & $64.55  \pm 0.03 $ & $23.8\pm0.1$ &  8.07 & S \\
mal32     & 2011 UB$_{411}$ & 55860.8 & $38.548\pm0.001$ & 25 & 2.0862 & 55804.0 & $39.601\pm 0.006$ & $0.0436 \pm0.0008 $ & $14.900\pm 0.001$ & $13.374 \pm  0.001$ & $322.5  \pm   0.6  $  & $50.471 \pm 0.014$ & $23.8\pm0.1$ &  7.92 & I \\
(Cookie)  &	            &         &		         &    &	       &	 &		     &		           &		       &		     &			     &		          &	         &       &   \\
mal12     & 2011 UP$_{410}$ & 55858.9 & $33.293\pm0.001$ & 32 & 2.2503 & 55810.0 & $39.602\pm 0.004$ & $0.1749 \pm0.0003 $ & $2.733 \pm 0.001$ & $241.740\pm  0.006$ & $179.2  \pm   0.2  $  & $339.745\pm 0.004$ & $23.4\pm0.1$ &  8.21 & S \\
\multicolumn{16}{c}{\textbf{Resonant 5:3} \citep{alexandersen14d}}\\
mal60     & 2011 UM$_{411}$ & 55858.9 & $48.408\pm0.007$ & 12 & 1.2891 & 55858.9 & $42.222\pm 0.014$ & $0.1467 \pm0.0004 $ & $9.385 \pm 0.003$ & $11.445 \pm  0.006$ & $197    \pm   4    $  & $183.8  \pm 0.2  $ & $24.1\pm0.1$ &  7.24 & S \\
mal34     & 2011 UL$_{411}$ & 55858.9 & $39.189\pm0.002$ & 16 & 1.4340 & 55803.0 & $42.224\pm 0.005$ & $0.07190\pm0.00009$ & $1.974 \pm 0.000$ & $308.53 \pm  0.04 $ & $82     \pm   2    $  & $358.747\pm 0.000$ & $24.3\pm0.1$ &  8.37 & S \\
mal14     & 2011 UO$_{411}$ & 55860.8 & $33.891\pm0.001$ & 26 & 1.4395 & 55804.0 & $42.282\pm 0.011$ & $0.2061 \pm0.0006 $ & $5.289 \pm 0.000$ & $310.100\pm  0.010$ & $98.9   \pm   0.5  $  & $347.583\pm 0.005$ & $23.8\pm0.1$ &  8.51 & S \\
mal27     & 2011 UJ$_{411}$ & 55858.9 & $37.500\pm0.002$ & 21 & 1.4338 & 55803.0 & $42.290\pm 0.009$ & $0.1189 \pm0.0007 $ & $5.141 \pm 0.001$ & $21.604 \pm  0.003$ & $30.0   \pm   1.0  $  & $344.336\pm 0.006$ & $23.9\pm0.1$ &  8.20 & S \\
mal15     & 2011 UK$_{411}$ & 55858.9 & $33.939\pm0.001$ & 27 & 1.4724 & 55803.0 & $42.32 \pm 0.02 $ & $0.2328 \pm0.0011 $ & $13.424\pm 0.001$ & $219.278\pm  0.001$ & $210.3  \pm   0.3  $  & $335.30 \pm 0.02 $ & $21.9\pm0.1$ &  6.66 & S \\
mal13     & 2011 UN$_{411}$ & 55860.8 & $33.447\pm0.002$ & 23 & 1.4395 & 55804.0 & $42.341\pm 0.007$ & $0.2104 \pm0.0003 $ & $3.234 \pm 0.000$ & $326.45 \pm  0.02 $ & $61.0   \pm   0.7  $  & $2.553  \pm 0.001$ & $24.2\pm0.2$ &  8.99 & S \\
\multicolumn{16}{c}{\textbf{Resonant 19:10} \citep{alexandersen14c}}\\
mal18     & 2011 UH$_{411}$ & 55858.9 & $35.636\pm0.002$ & 18 & 1.4504 & 55803.0 & $46.14 \pm 0.02 $ & $0.2386 \pm0.0009 $ & $27.041\pm 0.001$ & $214.057\pm  0.000$ & $198.3  \pm   0.5  $  & $346.696\pm 0.009$ & $24.2\pm0.1$ &  8.72 & I \\
\multicolumn{16}{c}{\textbf{Resonant 2:1} \citep{alexandersen14b}}\\
mal22     & 2011 UE$_{411}$ & 55858.9 & $36.574\pm0.002$ & 14 & 1.3792 & 55803.1 & $47.61 \pm 0.09 $ & $0.320  \pm0.003  $ & $9.355 \pm 0.001$ & $17.099 \pm  0.003$ & $73.3   \pm   0.3  $  & $329.11 \pm 0.09 $ & $24.6\pm0.1$ &  9.00 & S \\
mal47     & 2011 UG$_{411}$ & 55860.8 & $41.990\pm0.004$ & 18 & 1.2784 & 55859.8 & $47.66 \pm 0.04 $ & $0.139  \pm0.003  $ & $10.692\pm 0.001$ & $238.833\pm  0.004$ & $116    \pm   2    $  & $27.11  \pm 0.04 $ & $24.6\pm0.2$ &  8.35 & S \\
mal25     & 2011 UF$_{411}$ & 55860.8 & $37.267\pm0.002$ & 31 & 1.4313 & 55804.0 & $47.73 \pm 0.09 $ & $0.344  \pm0.002  $ & $5.872 \pm 0.001$ & $261.775\pm  0.010$ & $195.8  \pm   0.2  $  & $324.76 \pm 0.10 $ & $24.1\pm0.1$ &  8.35 & S \\
\multicolumn{16}{c}{\textbf{Resonant 5:2} \citep{alexandersen14g}}\\
mah04     & 2012 UJ$_{177}$ & 56221.9 & $34.429\pm0.001$ & 25 & 1.4585 & 55805.0 & $55.20 \pm 0.07 $ & $0.4333 \pm0.0013 $ & $15.632\pm 0.000$ & $298.867\pm  0.003$ & $46.9   \pm   0.2  $  & $17.41  \pm 0.04 $ & $24.1\pm0.1$ &  8.65 & I \\
mal61     & 2011 UT$_{411}$ & 55860.8 & $51.239\pm0.004$ & 20 & 1.4338 & 55803.1 & $55.7  \pm 0.3  $ & $0.406  \pm0.005  $ & $6.420 \pm 0.002$ & $4.876  \pm  0.011$ & $130.2  \pm   0.2  $  & $304.2  \pm 0.4  $ & $24.5\pm0.1$ &  7.46 & I \\
\multicolumn{16}{c}{\textbf{Resonant 3:1} \citep{alexandersen14f}}\\
mal23     & 2011 UR$_{411}$ & 55860.8 & $37.046\pm0.003$ & 15 & 1.2866 & 55859.9 & $62.1  \pm 0.2  $ & $0.438  \pm0.003  $ & $26.581\pm 0.002$ & $23.363 \pm  0.001$ & $45.8   \pm   0.6  $  & $346.99 \pm 0.07 $ & $24.0\pm0.1$ &  8.33 & I \\
mal08     & 2011 US$_{411}$ & 55860.8 & $31.226\pm0.002$ & 21 & 1.4308 & 55810.1 & $62.431\pm 0.011$ & $0.49983\pm0.00008$ & $22.040\pm 0.001$ & $24.813 \pm  0.000$ & $9.4    \pm   0.3  $  & $359.981\pm 0.000$ & $24.0\pm0.1$ &  9.11 & S \\
mal62     & 2011 UQ$_{411}$ & 55860.8 & $52.005\pm0.002$ & 25 & 1.4396 & 55804.0 & $62.4  \pm 0.2  $ & $0.405  \pm0.004  $ & $40.400\pm 0.002$ & $215.504\pm  0.000$ & $263.65 \pm   0.02 $  & $315.5  \pm 0.2  $ & $23.9\pm0.1$ &  6.76 & I \\
\multicolumn{16}{c}{\textbf{Resonant 4:1} \citep{alexandersen14e}}\\
mal33     & 2011 UP$_{411}$ & 55860.8 & $38.569\pm0.003$ & 17 & 1.4228 & 55810.1 & $75.79 \pm 0.02 $ & $0.49117\pm0.00015$ & $13.435\pm 0.001$ & $231.961\pm  0.002$ & $160.3  \pm   0.5  $  & $359.450\pm 0.000$ & $24.2\pm0.1$ &  8.35 & S \\
\multicolumn{16}{c}{\textbf{Resonant 16:3} \citep{alexandersen14h}}\\
mah08     & 2012 UK$_{177}$ & 56220.9 & $36.784\pm0.001$ & 25 & 1.4560 & 55805.9 & $92.3  \pm 0.2  $ & $0.6173 \pm0.0009 $ & $24.852\pm 0.001$ & $251.482\pm  0.002$ & $162.5  \pm   0.2  $  & $354.981\pm 0.014$ & $24.1\pm0.1$ &  8.38 & I \\
\multicolumn{16}{c}{\textbf{Inner classical} \citep{alexandersen14j}}\\
mal29     & 2011 UO$_{412}$ & 55860.8 & $37.892\pm0.002$ & 26 & 1.4146 & 55810.1 & $38.04 \pm 0.02 $ & $0.129  \pm0.003  $ & $27.787\pm 0.001$ & $21.128 \pm  0.000$ & $273.13 \pm   0.12 $  & $80.85  \pm 0.08 $ & $23.4\pm0.1$ &  7.62 & I \\
mal41     & 2011 UN$_{412}$ & 55858.9 & $40.960\pm0.004$ & 13 & 1.2892 & 55858.9 & $38.96 \pm 0.02 $ & $0.058  \pm0.003  $ & $20.411\pm 0.002$ & $216.340\pm  0.001$ & $326    \pm   5    $  & $209.1  \pm 0.2  $ & $24.5\pm0.1$ &  8.37 & S \\
\multicolumn{16}{c}{\textbf{Main classical} \citep{alexandersen14i}}\\
mah05     & 2012 UL$_{177}$ & 56220.9 & $36.474\pm0.001$ & 30 & 1.4643 & 55805.9 & $40.666\pm 0.009$ & $0.1166 \pm0.0008 $ & $19.236\pm 0.000$ & $272.168\pm  0.003$ & $147.6  \pm   0.6  $  & $335.255\pm 0.009$ & $22.9\pm0.1$ &  7.24 & S \\
mal24     & 2011 UK$_{412}$ & 55860.8 & $37.131\pm0.002$ & 29 & 1.4476 & 55804.0 & $40.737\pm 0.007$ & $0.0920 \pm0.0007 $ & $26.365\pm 0.001$ & $218.923\pm  0.001$ & $151.6  \pm   1.3  $  & $14.430 \pm 0.005$ & $23.4\pm0.1$ &  7.74 & S \\
mah13     & 2012 UO$_{177}$ & 56221.9 & $42.380\pm0.002$ & 25 & 1.4560 & 55805.9 & $41.562\pm 0.005$ & $0.035  \pm0.002  $ & $16.446\pm 0.000$ & $289.959\pm  0.006$ & $336    \pm   2    $  & $122.36 \pm 0.06 $ & $24.4\pm0.1$ &  8.13 & S \\
mal50     & 2011 UJ$_{412}$ & 55860.8 & $42.997\pm0.002$ & 17 & 1.4338 & 55803.1 & $42.031\pm 0.007$ & $0.029  \pm0.002  $ & $29.757\pm 0.002$ & $27.031 \pm  0.000$ & $148    \pm   5    $  & $219.82 \pm 0.10 $ & $24.3\pm0.2$ &  7.99 & S \\
mal37     & 2011 UX$_{411}$ & 55858.9 & $39.653\pm0.002$ & 23 & 1.3793 & 55803.1 & $42.43 \pm 0.02 $ & $0.098  \pm0.002  $ & $7.210 \pm 0.001$ & $24.163 \pm  0.002$ & $315.7  \pm   1.0  $  & $44.02  \pm 0.03 $ & $24.2\pm0.2$ &  8.27 & I \\
mal44     & 2011 UH$_{412}$ & 55860.8 & $41.372\pm0.002$ & 22 & 1.4313 & 55804.0 & $42.60 \pm 0.03 $ & $0.116  \pm0.003  $ & $15.118\pm 0.002$ & $21.517 \pm  0.001$ & $94.4   \pm   0.2  $  & $290.88 \pm 0.07 $ & $24.2\pm0.1$ &  8.02 & S \\
mah09     & 2012 UP$_{177}$ & 56221.9 & $39.922\pm0.002$ & 27 & 1.4585 & 55805.0 & $42.823\pm 0.014$ & $0.099  \pm0.002  $ & $16.601\pm 0.001$ & $280.170\pm  0.004$ & $59.9   \pm   0.8  $  & $42.49  \pm 0.02 $ & $24.0\pm0.1$ &  7.97 & S \\
mal49     & 2011 UL$_{412}$ & 55860.8 & $42.833\pm0.002$ & 21 & 1.4395 & 55804.0 & $43.03 \pm 0.02 $ & $0.083  \pm0.003  $ & $10.371\pm 0.002$ & $239.455\pm  0.005$ & $60.14  \pm   0.10 $  & $82.06  \pm 0.07 $ & $23.7\pm0.1$ &  7.42 & I \\
mal40     & 1999 RU$_{205}$ & 55858.9 & $40.875\pm0.002$ & 24 & 1.4339 & 55803.0 & $43.048\pm 0.012$ & $0.071  \pm0.002  $ & $7.737 \pm 0.001$ & $14.293 \pm  0.003$ & $329.4  \pm   1.4  $  & $41.43  \pm 0.02 $ & $23.1\pm0.1$ &  7.04 & S \\
mah07     & 2012 UN$_{177}$ & 56220.9 & $36.611\pm0.002$ & 29 & 1.4667 & 55805.0 & $43.181\pm 0.010$ & $0.1589 \pm0.0006 $ & $20.094\pm 0.001$ & $341.291\pm  0.002$ & $71.2   \pm   0.7  $  & $345.867\pm 0.006$ & $22.9\pm0.1$ &  7.28 & S \\
mal43     & 2011 UV$_{411}$ & 55858.9 & $41.355\pm0.002$ & 23 & 1.4340 & 55803.0 & $43.363\pm 0.013$ & $0.067  \pm0.002  $ & $4.419 \pm 0.001$ & $20.174 \pm  0.004$ & $323    \pm   2    $  & $43.37  \pm 0.03 $ & $24.1\pm0.1$ &  7.93 & S \\
mal39     & 2011 UC$_{412}$ & 55858.9 & $40.517\pm0.002$ & 26 & 1.4723 & 55803.0 & $43.408\pm 0.006$ & $0.0705 \pm0.0008 $ & $3.132 \pm 0.001$ & $337.47 \pm  0.02 $ & $33     \pm   2    $  & $17.794 \pm 0.007$ & $23.8\pm0.1$ &  7.75 & S \\
mal59     & 2011 UE$_{412}$ & 55858.9 & $48.245\pm0.003$ & 26 & 1.4339 & 55803.0 & $43.496\pm 0.005$ & $0.1119 \pm0.0005 $ & $22.799\pm 0.002$ & $218.117\pm  0.000$ & $5.6    \pm   1.4  $  & $165.93 \pm 0.06 $ & $23.3\pm0.1$ &  6.45 & S \\
mal54     & 2011 UB$_{412}$ & 55858.9 & $44.937\pm0.005$ & 17 & 1.2809 & 55858.9 & $43.50 \pm 0.02 $ & $0.056  \pm0.005  $ & $15.311\pm 0.002$ & $14.512 \pm  0.002$ & $246    \pm   4    $  & $123.94 \pm 0.14 $ & $24.3\pm0.1$ &  7.76 & S \\
mah12     & 2012 UM$_{177}$ & 56220.9 & $41.687\pm0.002$ & 19 & 1.4586 & 55805.0 & $43.549\pm 0.011$ & $0.058  \pm0.002  $ & $16.177\pm 0.000$ & $287.738\pm  0.006$ & $57     \pm   2    $  & $40.52  \pm 0.02 $ & $24.6\pm0.2$ &  8.33 & S \\
mal46     & 2011 UW$_{411}$ & 55858.9 & $41.526\pm0.002$ & 20 & 1.4340 & 55803.0 & $43.84 \pm 0.04 $ & $0.165  \pm0.003  $ & $4.529 \pm 0.002$ & $18.484 \pm  0.005$ & $94.7   \pm   0.2  $  & $297.62 \pm 0.08 $ & $24.5\pm0.2$ &  8.37 & I \\
mal52     & 2011 UU$_{411}$ & 55858.9 & $43.301\pm0.002$ & 24 & 1.4723 & 55803.0 & $43.941\pm 0.008$ & $0.031  \pm0.002  $ & $1.694 \pm 0.001$ & $357.00 \pm  0.03 $ & $99     \pm   2    $  & $299.72 \pm 0.03 $ & $24.3\pm0.1$ &  7.96 & S \\
mal42     & 2001 RZ$_{143}$ & 55858.9 & $41.288\pm0.002$ & 23 & 1.3795 & 55803.0 & $44.061\pm 0.005$ & $0.0640 \pm0.0005 $ & $2.122 \pm 0.002$ & $8.32   \pm  0.02 $ & $35     \pm   3    $  & $350.411\pm 0.004$ & $22.6\pm0.1$ &  6.44 & S \\
mal51     & 2011 UY$_{411}$ & 55858.9 & $43.191\pm0.002$ & 25 & 1.4423 & 55803.0 & $44.22 \pm 0.02 $ & $0.064  \pm0.002  $ & $1.907 \pm 0.000$ & $309.43 \pm  0.04 $ & $154.1  \pm   0.7  $  & $294.69 \pm 0.04 $ & $24.0\pm0.1$ &  7.64 & S \\
mal56     & 2011 UF$_{412}$ & 55858.9 & $45.576\pm0.004$ & 12 & 1.1466 & 55803.1 & $45.07 \pm 0.02 $ & $0.012  \pm0.003  $ & $2.146 \pm 0.000$ & $313.97 \pm  0.07 $ & $278    \pm  41    $  & $160.7  \pm 0.9  $ & $24.4\pm0.1$ &  7.83 & S \\
mal38     & 2011 UM$_{412}$ & 55860.8 & $40.339\pm0.002$ & 27 & 1.4309 & 55804.1 & $45.38 \pm 0.02 $ & $0.132  \pm0.002  $ & $10.729\pm 0.001$ & $231.692\pm  0.002$ & $124.1  \pm   1.0  $  & $28.50  \pm 0.02 $ & $24.4\pm0.2$ &  8.36 & I \\
mal53     & 2011 UD$_{412}$ & 55858.9 & $44.398\pm0.001$ & 22 & 2.0893 & 55803.0 & $45.419\pm 0.006$ & $0.0535 \pm0.0011 $ & $5.496 \pm 0.001$ & $358.349\pm  0.007$ & $101.0  \pm   0.5  $  & $297.64 \pm 0.01 $ & $23.2\pm0.1$ &  6.75 & S \\
mal55     & 2011 UZ$_{411}$ & 55858.9 & $45.158\pm0.002$ & 22 & 1.4504 & 55803.0 & $45.535\pm 0.014$ & $0.053  \pm0.003  $ & $1.214 \pm 0.000$ & $315.36 \pm  0.07 $ & $352.2  \pm   0.3  $  & $77.97  \pm 0.04 $ & $24.3\pm0.1$ &  7.75 & S \\
mal36     & 2011 UA$_{412}$ & 55858.9 & $39.507\pm0.003$ & 12 & 1.1468 & 55803.0 & $46.250\pm 0.011$ & $0.1476 \pm0.0008 $ & $2.956 \pm 0.001$ & $252.56 \pm  0.02 $ & $146    \pm   2    $  & $352.267\pm 0.004$ & $24.6\pm0.2$ &  8.66 & I \\
mal48     & 2011 UG$_{412}$ & 55860.8 & $42.436\pm0.002$ & 33 & 1.4313 & 55804.0 & $46.45 \pm 0.03 $ & $0.155  \pm0.002  $ & $2.496 \pm 0.000$ & $303.96 \pm  0.03 $ & $24.9   \pm   0.4  $  & $48.71  \pm 0.06 $ & $23.8\pm0.1$ &  7.57 & S \\
\multicolumn{16}{c}{\textbf{Outer classical} \citep{alexandersen14l}}\\
mal58     & 2011 US$_{412}$ & 55858.9 & $46.661\pm0.004$ & 16 & 1.4501 & 55803.1 & $47.88 \pm 0.08 $ & $0.161  \pm0.005  $ & $2.606 \pm 0.002$ & $5.09   \pm  0.02 $ & $116.38 \pm   0.04 $  & $288.2  \pm 0.2  $ & $24.4\pm0.1$ &  7.74 & S \\
mal57     & 2011 UT$_{412}$ & 55860.8 & $46.124\pm0.003$ & 15 & 1.4395 & 55804.0 & $48.08 \pm 0.07 $ & $0.183  \pm0.004  $ & $17.822\pm 0.002$ & $224.314\pm  0.002$ & $79.71  \pm   0.05 $  & $66.9   \pm 0.2  $ & $24.1\pm0.1$ &  7.50 & I \\
\multicolumn{16}{c}{\textbf{Detached} \citep{alexandersen14k}}\\
mal20     & 2011 UR$_{412}$ & 55860.8 & $36.158\pm0.002$ & 20 & 1.4395 & 55804.0 & $50.07 \pm 0.06 $ & $0.314  \pm0.002  $ & $17.873\pm 0.001$ & $15.280 \pm  0.001$ & $54.2   \pm   0.4  $  & $340.63 \pm 0.03 $ & $24.2\pm0.2$ &  8.60 & S \\
mah10     & 2012 UQ$_{177}$ & 56220.9 & $40.257\pm0.002$ & 26 & 1.4560 & 55805.9 & $51.97 \pm 0.07 $ & $0.312  \pm0.002  $ & $19.603\pm 0.001$ & $330.160\pm  0.003$ & $120.1  \pm   0.2  $  & $328.58 \pm 0.07 $ & $23.8\pm0.1$ &  7.77 & S \\
mah06     & 2012 US$_{177}$ & 56221.9 & $36.605\pm0.002$ & 25 & 1.4665 & 55805.1 & $56.09 \pm 0.05 $ & $0.3665 \pm0.0011 $ & $17.214\pm 0.001$ & $271.782\pm  0.003$ & $92.5   \pm   0.3  $  & $11.89  \pm 0.02 $ & $24.2\pm0.1$ &  8.51 & S \\
mal26     & 2011 UP$_{412}$ & 55858.9 & $37.450\pm0.002$ & 23 & 1.4422 & 55803.0 & $56.34 \pm 0.03 $ & $0.3443 \pm0.0007 $ & $19.721\pm 0.001$ & $23.081 \pm  0.000$ & $26.4   \pm   0.4  $  & $351.390\pm 0.008$ & $23.8\pm0.1$ &  8.09 & S \\
mal19     & 2011 UQ$_{412}$ & 55858.9 & $35.912\pm0.001$ & 22 & 1.4311 & 55803.0 & $67.32 \pm 0.08 $ & $0.4829 \pm0.0009 $ & $16.712\pm 0.001$ & $219.944\pm  0.000$ & $197.5  \pm   0.2  $  & $352.205\pm 0.014$ & $23.0\pm0.1$ &  7.51 & S \\
mah14     & 2012 UR$_{177}$ & 56221.9 & $44.388\pm0.003$ & 26 & 1.4560 & 55805.9 & $73.8  \pm 0.3  $ & $0.492  \pm0.003  $ & $16.353\pm 0.000$ & $291.500\pm  0.006$ & $158.2  \pm   0.2  $  & $340.65 \pm 0.13 $ & $24.2\pm0.1$ &  7.76 & I \\
\multicolumn{16}{c}{\textbf{Unclassified} \citep{tomatic14a}}\\
mah11nt   &                 & 56221.9 & $38    \pm4    $ &  3 & 0.0002 & 56221.9 & $39    \pm21    $ & $0.0    \pm0.6    $ & $15    \pm 4    $ & $316    \pm 80    $ & $88     \pm1384    $  & $352    \pm 7    $ & $24.2\pm0.1$ &  8.30 & N \\
mal11nt   & 2011 UU$_{412}$ & 55860.8 & $32.8  \pm0.5  $ & 11 & 0.0986 & 55859.8 & $39    \pm10    $ & $0.2    \pm0.5    $ & $23    \pm 2    $ & $25.5   \pm  0.7  $ & $32     \pm 248    $  & $343    \pm 7    $ & $24.5\pm0.1$ &  9.40 & N \\
\multicolumn{16}{c}{\textbf{Uncharacterised \& unclassified} \citep{alexandersen14m,tomatic14a}}\\
umal64nt  &                 & 55858.9 & $36    \pm4    $ &  3 & 0.0002 & 55858.9 & $37    \pm20    $ & $0.0    \pm0.6    $ & $3     \pm 3    $ & $291    \pm277    $ & $98     \pm1309    $  & $0.8    \pm 0.7  $ & $24.7\pm0.3$ &  9.09 & N \\
umal70nt  & 2011 UV$_{412}$ & 55858.9 & $47.85 \pm0.05 $ &  9 & 0.8354 & 55858.9 & $48    \pm 5    $ & $0.3    \pm0.2    $ & $11.54 \pm 0.03 $ & $219.29 \pm  0.03 $ & $63     \pm  11    $  & $74     \pm12    $ & $24.7\pm0.1$ &  7.89 & N \\
umal65nt  & 2011 UY$_{412}$ & 55858.9 & $42    \pm2    $ &  4 & 0.0195 & 55858.9 & $43    \pm22    $ & $0.0    \pm0.6    $ & $8     \pm 3    $ & $16     \pm  6    $ & $20     \pm1486    $  & $357    \pm 3    $ & $24.7\pm0.1$ &  8.52 & N \\
umah16nt  & 2012 UU$_{177}$ & 56221.9 & $45    \pm2    $ &  7 & 0.0055 & 56220.0 & $46    \pm24    $ & $0.0    \pm0.6    $ & $16.1  \pm 0.6  $ & $284    \pm  5    $ & $119    \pm1789    $  & $352    \pm 8    $ & $24.9\pm0.3$ &  8.30 & N \\
umal66nt  &                 & 55858.9 & $46    \pm5    $ &  3 & 0.0002 & 55858.9 & $47    \pm24    $ & $0.0    \pm0.6    $ & $13    \pm25    $ & $24     \pm 19    $ & $13     \pm1620    $  & $357    \pm 3    $ & $24.6\pm0.1$ &  8.06 & N \\
umah17    & 2012 UT$_{177}$ & 56221.9 & $52.65 \pm0.01 $ & 21 & 1.4614 & 55806.0 & $47.49 \pm 0.08 $ & $0.127  \pm0.007  $ & $16.363\pm 0.001$ & $320.497\pm  0.013$ & $282    \pm   6    $  & $144.9  \pm 0.5  $ & $24.8\pm0.1$ &  7.49 & N \\
umal68nt  &                 & 55858.9 & $47    \pm5    $ &  3 & 0.0002 & 55858.9 & $48    \pm25    $ & $0.0    \pm0.6    $ & $8     \pm24    $ & $10     \pm 60    $ & $19     \pm1680    $  & $0.7    \pm 0.7  $ & $24.8\pm0.1$ &  8.04 & N \\
umal69nt  &                 & 55858.9 & $47    \pm5    $ &  3 & 0.0002 & 55858.9 & $48    \pm25    $ & $0.0    \pm0.6    $ & $3     \pm20    $ & $355    \pm304    $ & $40     \pm1708    $  & $358    \pm 2    $ & $24.8\pm0.1$ &  8.10 & N \\
umal63    & 2011 UW$_{412}$ & 55860.8 & $38.419\pm0.003$ & 18 & 1.4228 & 55810.1 & $78.2  \pm 0.2  $ & $0.521  \pm0.002  $ & $12.976\pm 0.001$ & $233.178\pm  0.002$ & $134.1  \pm   0.5  $  & $6.17   \pm 0.03 $ & $24.7\pm0.2$ &  8.86 & N \\
umal67nt  & 2011 UX$_{412}$ & 55858.9 & $46.9  \pm0.4  $ &  7 & 0.1010 & 55858.9 & $54    \pm25    $ & $0.4    \pm0.4    $ & $29    \pm 2    $ & $216.9  \pm  0.4  $ & $270    \pm  16    $  & $311    \pm34    $ & $24.6\pm0.2$ &  7.94 & N \\
\enddata
\end{deluxetable}

\clearpage

\begin{deluxetable}{cc|c|c|c|c|c}
\tabletypesize{\scriptsize}
\tablecaption{Population estimates for stable Neptunian Trojans, temporary Neptunian co-orbitals and temporary Uranian co-orbitals. 
Population estimates are medians given with $95\%$ confidence ranges and upper limits are $95\%$ confidence upper limits. 
For Trojans, co-orbitals and Plutinos, estimates are given for two different magnitude-distribution models, a knee with $H_t=7.7$ to $\alpha_f=0.40$ and a divot with $H_t=8.4$, $c=1$ to $\alpha_f=0.80$. Both models have $\alpha_b=0.95$.
For the 3:1 and 4:1 resonance, a single exponential with $\alpha=0.9$ was used. 
\label{tab:popest}}
\tablehead{
\colhead{}                              & \colhead{} & \colhead{$N(H_r\le8.4)$} & \colhead{$N(H_r\le8.66)$} & \colhead{$N(H_r\le9.1)$} & \colhead{$N(H_r\le10.0)$} & \colhead{$N(H_r\le11.0)$} \\
}
\startdata
\multirow{2}{*}{Stable Neptunian Trojans}        & Knee  & $\le250$               & $\le250$                & $80^{+300}_{-70}$       & $140^{+600}_{-130}$       &                        \\
                                                 & Divot & $\le250$               & $\le260$                & $80^{+300}_{-70}$       & $150^{+600}_{-140}$       &                        \\
\tableline                                                                                                                   
\multirow{2}{*}{Temporary Neptunian co-orbitals} & Knee  & $70^{+300}_{-60}$      & $70^{+300}_{-60}$       & $90^{+400}_{-80}$       & $200^{+900}_{-190}$       & $1200^{+3000}_{-1000}$ \\
                                                 & Divot & $70^{+300}_{-60}$      & $80^{+300}_{-70}$       & $90^{+400}_{-80}$       & $210^{+900}_{-200}$       & $2500^{+6000}_{-2100}$ \\
\tableline                                                                                                                        
\multirow{2}{*}{Temporary Uranian co-orbitals}   & Knee  & $\le300$               & $\le300$                & $\le300$                & $110^{+400}_{-100}$       & $190^{ +800}_{-180}$   \\
                                                 & Divot & $\le300$               & $\le300$                & $\le300$                & $110^{+500}_{-100}$       & $270^{+1200}_{-260}$   \\
\tableline                                                                                                                                   
\multirow{2}{*}{Plutinos}                        & Knee  & $7000^{+3000}_{-2000}$ & $9000\pm3000$           & $14000^{+5000}_{-4000}$ & $35000^{+12000}_{-10000}$ &                        \\
                                                 & Divot & $8000^{+3000}_{-2000}$ & $9000\pm3000$           & $12000^{+4000}_{-3000}$ & $37000^{+12000}_{-10000}$ &                        \\
\tableline                                                                                                                                   
3:1 ($q>31.2$, $e>0.40$)                         & Exponential  &                 & $3000^{+4000}_{-2000}$  &                         &                           &                        \\
\tableline                                                                                                                                   
4:1 ($q>31.2$, $e>0.49$)                         & Exponential  &                 & $2500^{+10000}_{-2400}$ &                         &                           &                        \\
\enddata
\end{deluxetable}

\clearpage

\end{document}